\documentclass[aps,pre,twocolumn,superscriptaddress]{revtex4}

\usepackage{bm}
\usepackage{mathrsfs}
\usepackage{graphicx}
\usepackage{color}
\usepackage{amsmath}
\usepackage{amssymb}

\begin{document}
 
\title{Brownian yet non-Gaussian diffusion: from superstatistics to subordination
of diffusing diffusivities}

\author{Aleksei V. Chechkin}
\affiliation{Institute for Physics \& Astronomy, University of Potsdam, 14476
Potsdam-Golm, Germany}
\affiliation{INFN, Padova Section, and Department of Physics and Astronomy "G.
Galilei", University of Padova, Via Marzolo 8 35131, Padova, Italy}
\affiliation{Akhiezer Institute for Theoretical Physics, Kharkov 61108, Ukraine}
\author{Flavio Seno}
\affiliation{INFN, Padova Section, and Department of Physics and Astronomy "G.
Galilei", University of Padova, Via Marzolo 8 35131, Padova, Italy}
\author{Ralf Metzler}
\affiliation{Institute for Physics \& Astronomy, University of Potsdam, 14476
Potsdam-Golm, Germany}
\author{Igor M. Sokolov}
\affiliation{Institute of Physics, Humboldt University Berlin, Newtonstrasse 15,
D-12489 Berlin, Germany}
\date{\today}

\begin{abstract}
A growing number of biological, soft, and active matter systems are observed to
exhibit normal diffusive dynamics with a linear growth of the mean squared
displacement, yet with a non-Gaussian distribution of increments. Based on the
Chubinsky-Slater idea of a diffusing diffusivity we here establish and analyze
a minimal model framework of diffusion processes with fluctuating
diffusivity. In particular, we demonstrate the equivalence of the diffusing
diffusivity process with a superstatistical approach with a distribution of
diffusivities, at times shorter than the diffusivity correlation time. At
longer times a crossover to a Gaussian distribution with an effective
diffusivity emerges. Specifically, we establish a subordination picture of
Brownian but non-Gaussian diffusion processes, that can be used for a wide
class of diffusivity fluctuation statistics. Our results are shown to be in
excellent agreement with simulations and numerical evaluations.
\end{abstract}

\maketitle

\section{Introduction}

Thermally driven diffusive motion belongs to the fundamental physical processes.
To a big extent inspired by the groundbreaking experiments of Robert Brown in
the 1820ies \cite{brown} the theoretical foundations of the theory of diffusion
were then laid by Einstein, Sutherland, Smoluchowski, and Langevin between 1905
and 1908 \cite{einstein,sutherland,smoluchowski,langevin}. On their basis novel
experiments, such as the seminal works by Perrin and Nordlund \cite{perrin,
nordlund}, in turn delivered ever better quantitative information on molecular
diffusion as well as the atomistic nature of matter. Typically, we now identify
two fundamental properties with Brownian diffusive processes: (i) the linear growth
in time of the mean squared displacement (MSD)
\begin{equation}
\label{msd}
\langle\mathbf{r}^2(t)\rangle=\int_{-\infty}^{\infty}\mathbf{r}^2P(\mathbf{r},t)
d\mathbf{r}=2dDt
\end{equation}
typically termed normal (Fickian) diffusion. Here $d$ denotes the spatial dimension
and $D$
is called the diffusion coefficient. (ii) The second property is the Gaussian shape
\begin{equation}
\label{gauss}
P(\mathbf{r},t)=\frac{1}{(4\pi Dt)^{d/2}}\exp\left(-\frac{\mathbf{r}^2}{4Dt}\right)
\end{equation}
of the probability density function to find the diffusing particle at position
$\mathbf{r}$ at some time $t$ \cite{vankampen}. From a more mathematical
viewpoint the Gaussian emerges as limit distribution of independent,
identically distributed random variables (the steps of the random walk) with
finite variance and in that sense assumes a universal character \cite{gnedenko}.

Deviations from the linear time dependence (\ref{msd}) are routinely observed. Thus,
modern microscopic techniques reveal \emph{anomalous diffusion\/} with the power-law
dependence $\langle\mathbf{r}^2(t)\rangle\simeq t^{\alpha}$ of the MSD, where
according to
the value of the anomalous diffusion exponent we distinguish subdiffusion for $0
<\alpha<1$ and superdiffusion with $1<\alpha<2$ \cite{bouchaud,report,
pccp,bba,hoefling}. Examples for subdiffusion of passive molecular and submicron
tracers abound in the cytoplasm of living biological cells \cite{golding,goldinga,
goldingb} and in artificially crowded fluids \cite{banks}, as well as in quasi
two-dimensional systems such as lipid bilayer membranes
\cite{membranes,membranesa,membranesb,membranesc}. Superdiffusion is typically
associated with active processes and also observed in living cells \cite{elbaum}.
Anomalous diffusion
processes arise due to the loss of independence of the random variables, divergence
of the variance of the step length or the mean of the step time distribution,
as well as due to the tortuosity of the embedding space. The associated probability
density function of anomalous diffusion processes may have both Gaussian and
non-Gaussian shapes \cite{bouchaud,report,pccp}.

A new class of diffusive dynamics has recently been reported in a number of soft
matter, biological and other complex systems: in these processes the MSD is
normal of the form (\ref{msd}), however, the probability density function $P(
\mathbf{r},t)$ is non-Gaussian, typically characterized by a distinct exponential
shape
\begin{equation}
\label{expo}
P(\mathbf{r},t)\simeq\exp\left(-\frac{|\mathbf{r}|}{\lambda(t)}
\right),
\end{equation}
with the decay length $\lambda(t)=\sqrt{Dt}$ \cite{granick}.
This form of the probability density function is also sometimes
called a Laplace distribution. The Brownian yet non-Gaussian feature appears quite
robustly in a large range of systems, including beads diffusing on lipid tubes
\cite{wang} or in networks \cite{wang,gels}, tracer motion in colloidal, polymeric,
or active suspensions \cite{susp}, in biological cells \cite{bio}, as well as the
motion of individuals in heterogeneous populations such as nematodes \cite{hapca}.
For additional examples see \cite{bng,bnga,bngb,bngc} and the references in \cite{granick,gary,
klsebastian}.

How can this combination of normal, Brownian scaling of the mean squared displacement
be reconciled with the existence of a non-Gaussian probability density function? One
argument not brought forth in the discussion of anomalous diffusion above is the
possibility that the random variables making up the observed dynamics are indeed
not identically distributed. This fact can be introduced in different ways. First,
Granick and co-workers \cite{wang} as well as Hapca et al. \cite{hapca} employed
distributions of the diffusivity of individual tracer particles to explain this
remarkable behavior: indeed, averaging the Gaussian probability density function
(\ref{gauss}) for a single diffusivity $D$ over the exponential distribution $p_D(D)
=\langle D\rangle^{-1}\exp(-D/\langle D\rangle)$ with the mean diffusivity $\langle
D\rangle$, the exponential form (\ref{expo}) of the probability density function
emerges \cite{gary,hapca}. In fact, this idea of creating an ensemble behavior in
terms of distributions of diffusivities of individual tracer particles is analogous
to the concept of superstatistical Brownian motion: based on two statistical levels
describing, respectively, the fast jiggly dynamics of the Brownian particle and the
slow environmental fluctuations with spatially local patches of given diffusivity
this concept demonstrates how non-Gaussian probability densities arise physically
\cite{beck}. In what follows we refer to averaging over a diffusivity distribution
$p_D(D)$ as superstatistical approach. An important additional observation from
experiments that cannot be explained by the superstatistical approach is that
``the distribution function will converge to a Gaussian at times greater than the
correlation time of the fluctuations'' \cite{granick}. This is impressively
demonstrated, for instance, in Fig.~1C in \cite{wang}. This crossover cannot be
explained by the superstatistical approach. At the same time the normal-diffusive
behavior is not affected by the crossover between the shapes of the distribution.

Second, Chubinsky and Slater came up with the diffusing diffusivity model, in
which the diffusion coefficient of the tracer particle evolves in time like the
coordinate of a Brownian particle in a gravitational field \cite{gary}. For short
times they indeed find an exponential form (\ref{expo}). At long times, they
demonstrate from simulations that the probability density function crosses over to
a Gaussian shape. Jan and Sebastian formalize the diffusing diffusivity model in
an elegant path integral approach, which they explictly solve in two spatial
dimensions \cite{klsebastian}. Their results are consistent with those of
Ref.~\cite{gary}.

Here we introduce a simple yet powerful minimal model for diffusing diffusivities,
based on the concept of subordination. Based on a double Langevin equation approach
our model is fully analytical, providing an explicit solution for the probability
density function in Fourier space. The inversion is easily feasible numerically,
and we demonstrate excellent agreement with simulations of the underlying stochastic
equations. Moreover, we provide the analytical expressions for the asymptotic
behavior at short and long times, including the crossover to Gaussian statistics,
and derive explicit results for the kurtosis of
the probability density function. The bivariate Fokker-Planck equation for this
process and its connection to the subordination concept are established. Finally,
we show that at times shorter than the diffusivity correlation time our analytical
results are fully consistent with the superstatistical approach. Our approach has
the distinct advantage that it is amenable to a large variety of different
fluctuating diffusion scenarios.

In what follows we first formulate the coupled Langevin equations for the diffusing
diffusivity model. Section 3 then introduces the subordination concept allowing
us to derive the exact form of the subordinator as well as the Fourier image of
the probability density function. The Brownian form of the MSD is demonstrated
and the short and long time limits derived. Moreover, the connection to the
superstatistical approach is made. The kurtosis quantifying the non-Gaussian
shape of the probability density function is derived. In section 4 the
bivariate Fokker-Planck equation for the joint probability density function
$P(x,D,t)$ is analyzed, before drawing our conclusions in section 5. Several
Appendices provide additional details.

\section{Superstatistical approach to Brownian yet non-Gaussian diffusion}
\label{superstat}

As mentioned above, it was suggested by Granick and coworkers \cite{granick} as
well as by Hapca et al. \cite{hapca} that the Laplace distribution
\begin{equation}
\label{super0}
P(x,t)=\frac{1}{\sqrt{4\langle D\rangle t}}\exp\left(-\frac{|x|}{(\langle D\rangle
t)^{1/2}}\right)
\end{equation}
with effective diffusivity $\langle D\rangle$ emerging from a standard Gaussian
distribution
\begin{equation}
\label{gauss0}
G(x,t|D)=\frac{1}{\sqrt{4\pi Dt}}\exp\left(-\frac{x^2}{4Dt}\right)
\end{equation}
with diffusivity $D$, through the averaging procedure
\begin{equation}
\label{superdef}
P(x,t)=\int_0^{\infty}p_D(D)G(x,t|D)dD
\end{equation}
over $D$. This approach corresponds to the idea of superstatistics \cite{beck}:
accordingly the overall distribution function $P(x,t)$ of a system of tracer
particles, individually moving in sufficiently large, disjunct patches with local
diffusivity $D$, becomes the weighted average, where $p_D(D)$ is the stationary
state probability density for the particle diffusivities $D$. While in this Section
we restrict the discussion to the one-dimensional case, we will also provide
results for higher dimensions below.

Fourier transforming Eq.~(\ref{superdef}) we obtain
\begin{equation}
\label{fouriersuper}
P(k,t)=\int_0^{\infty}p_D(D)e^{-Dk^2t}dD=\tilde{p}_D(s=k^2t),
\end{equation}
where we used the fact that $G(k,t)=\exp(-Dk^2t)$. On the right hand side we
identified the integral of $p_D(D)$ over $\exp(-Dk^2t)$ as the Laplace transform
$\tilde{p}_D(s=k^2t)$ to be taken at $s=k^2t$. Concurrently, the Fourier
transform of expression (\ref{super0}) is
\begin{equation}
P(k,t)=\frac{1}{1+\langle D\rangle k^2t}.
\end{equation}
Combining these results and recalling the Laplace transform $\mathscr{L}\{\tau_{
\star}^{-1}\exp(-t/\tau_{\star})\}=(1+s\tau_{\star})$, we uniquely find that indeed
\begin{equation}
\label{expo0}
p_D(D)=\frac{1}{\langle D\rangle}\exp\left(-\frac{D}{\langle D\rangle}\right).
\end{equation}
To obtain the Laplace distribution (\ref{super0}) as superstatistical average of
elementary Gaussians (\ref{gauss0}), the necessary distribution of the diffusivities
is the exponential (\ref{expo0}). This is exactly the result of Granick and coworkers
\cite{granick} and Hapca et al.~\cite{hapca}. (We note that Hapca and coworkers also
report results for the case of a gamma distribution $p_D(D)$.)

Now, let us take the Fourier inversion of Eq.~(\ref{fouriersuper}) and invoke the
substitution $\kappa=kt^{1/2}$,
\begin{eqnarray}
\nonumber
P(x,t)&=&\frac{1}{2\pi}\int_{-\infty}^{\infty}e^{-ikx}\tilde{p}_D(k^2t)dk\\
&=&\frac{1}{
2\pi t^{1/2}}\int_{-\infty}^{\infty}e^{-i\kappa x/t^{1/2}}\tilde{p}_D(\kappa^2)d
\kappa.
\end{eqnarray}
The right hand side defines a scaling function $F$ of the form
\begin{equation}
P(x,t)=\frac{1}{t^{1/2}}F(\zeta),
\end{equation}
where $\zeta=x/t^{1/2}$. Thus the form $F$ as function of the similarity variable
$\zeta$ is an invariant. In particular, no transition of $P(x,t)$ from a Laplace
distribution to a different shape is possible in this superstatistical framework.
To account for the experimental observation, however, we are seeking a model to
explain the crossover from an initial Laplace distribution to a Gaussian shape at
long(er) times.

\subsection*{Anomalous diffusion with exponential, stretched Gaussian, and power
law shapes of the probability density function}

We briefly digress to mention that for the case of anomalous diffusion with a mean
squared displacement of the form $\langle x^2(t)\rangle\simeq t^{\alpha}$ a similar
phenomena was observed. Namely, for the motion of particles in a viscoelastic
environment with a fixed generalized diffusivity $D_{\alpha}$ of dimension $\mathrm{
cm}^2/\mathrm{sec}^{\alpha}$ the motion is characterized by the Gaussian
\cite{goychuk,pccp}
\begin{equation}
\label{pdfalpha}
G_{\alpha}(x,t|D_{\alpha})=\frac{1}{\sqrt{4\pi D_{\alpha}t^{\alpha}}}\exp\left(
-\frac{x^2}{4D_{\alpha}t^{\alpha}}\right)
\end{equation}
with $x/t^{\alpha/2}$ scaling variable. Instead, in a recent experimental study
observing the motion of labeled messenger RNA molecules in living E.coli and
S.cerevisiae cells an exponential distribution of the diffusivity was found
\cite{spako},
\begin{equation}
\label{pdalpha}
p_D(D_{\alpha})=\frac{1}{D_{\alpha}^{\star}}\exp\left(-\frac{D_{\alpha}}{D_{\alpha}
^{\star}}\right)
\end{equation}
on the single trajectory level, pointing at a higher inhomogeneity of the motion
than previously assumed. The distribution (\ref{pdalpha}) combined with the Gaussian
(\ref{pdfalpha}) gives rise to the Laplace distribution \cite{spako}
\begin{equation}
P_{\alpha}(x,t)=\frac{1}{\sqrt{4D_{\alpha}^{\star}t^{\alpha}}}\exp\left(-
\frac{|x|}{\sqrt{D_{\alpha}^{\star}t^{\alpha}}}\right).
\end{equation}

Similarly one can show that the stretched Gaussian observed for the lipid motion
in protein-crowded lipid bilayer membranes \cite{membranesa} emerges from the
Gaussian (\ref{pdfalpha}) in terms of a modified diffusivity distribution of the form
\begin{equation}
p_D(D_{\alpha})=\frac{1}{\Gamma(1+1/\kappa)D_{\alpha}^{\star}}\exp\left(-
\left[\frac{D_{\alpha}}{D_{\alpha}^{\star}}\right]^{\kappa}\right).
\end{equation}
In that case the resulting distribution assumes the form
\begin{equation}
P_{\alpha}(x,t)\simeq\exp\left(-c\left[\frac{|x|}{(4D_{\alpha}^{\star}t^{\alpha})^{
1/2}}\right]^{2\kappa/(1+\kappa)}\right)
\end{equation}
with an additional power law term in $x$, see Appendix \ref{appc}. Depending on the
value of $\kappa$ one can then obtain stretched Gaussian shapes for $p_{\alpha}(x,t)$
or even broader than exponential forms (superstretched Gaussians).

We finally note that for a power law distribution
\begin{equation}
p_D(D)\simeq D^{-1-\alpha}
\end{equation}
with $0<\alpha<2$ the resulting superstatistical distribution acquires long tails
of the form
\begin{equation}
P_{\alpha}(x,t)\simeq\frac{1}{|x|^{2\alpha+1}},
\end{equation}
as demonstrated in Appendix \ref{appc}. This brief discussion shows the need for
a more general model for the diffusing diffusivity, the basis of which is
established here.

\section{Langevin model for diffusing diffusivities}

To describe Brownian but non-Gaussian diffusion we start with the combined set of
stochastic equations
\begin{subequations}
\label{lang}
\begin{eqnarray}
\label{lang1}
\frac{d}{dt}\mathbf{r}(t)&=&\sqrt{2D(t)}\bm{\xi}(t),\\
\label{lang2}
D(t)&=&\mathbf{Y}^2(t),\\
\label{lang3}
\frac{d}{dt}\mathbf{Y}(t)&=&-\frac{1}{\tau}\mathbf{Y}+\sigma\bm{\eta}(t).
\end{eqnarray}
The independent noise terms $\bm{\xi}(t)$ and $\bm{\eta}(t)$ are white and
Gaussian, and both are specified by their first two moments
\begin{eqnarray}
&&\hspace*{-0.8cm}
\langle\bm{\xi}(t)\rangle=0,\quad\langle\xi_i(t_1)\xi_j(t_2)\rangle=\delta_{
ij}\delta(t_1-t_2)\\
&&\hspace*{-0.8cm}
\langle\bm{\eta}(t)\rangle=0,\quad\langle\eta_l(t_1)\eta_m(t_2)\rangle=\delta_{
lm}\delta(t_1-t_2),
\end{eqnarray}
for $i,j=x,y,z$ and $l,m=1,\ldots,n$. As explained below, the dimension $n$ of
the process $\mathbf{Y}(t)$ may differ from the value of $d$ of the process
$\mathbf{r}(t)$ in real space.

In the above set of coupled stochastic equations (\ref{lang}),
expression (\ref{lang1}) designates the well-known overdamped Langevin equation
driven by the white Gaussian noise $\bm{\xi}(t)$ \cite{risken}. However, we
consider the diffusion coefficient $D(t)$ to be a random function of time, and
we express it in terms of the square of the Ornstein-Uhlenbeck process $\mathbf{
Y}(t)$ (see below for the reasoning).
The physical dimension of the latter is $[\mathbf{Y}]=\mathrm{
cm}/\mathrm{sec}^{1/2}$. In Eq.~(\ref{lang3}) the correlation time of the
Ornstein-Uhlenbeck process is $\tau$, and $\sigma$ of units $[\sigma]=\mathrm{cm}/
\mathrm{sec}$ characterizes the amplitude of the fluctuations of $\mathbf{Y}$.
We complete the set of stochastic equations with the initial
conditions, chosen as
\begin{equation}
\mathbf{r}(0)=0,\quad \mathbf{Y}(0)=\mathbf{Y}_0.
\end{equation}
\end{subequations}

Physically, the choice of the above set of dynamic equations corresponds to the
following reasonings. In the diffusing diffusivity picture we model the particle
motion, on the single trajectory level, by the random diffusivity $D(t)$. Taking
$D(t)$ as the square of the auxiliary variable $\mathbf{Y}(t)$ guarantees the
non-negativity of $D(t)$. This way we avoid the need to impose reflecting boundary
condition on $D(t)$ at $D=0$, which is more difficult to handle analytically
\cite{gary}. The reason to choose the Ornstein-Uhlenbeck process (\ref{lang3}) for
$\mathbf{Y}(t)$ is two-fold. First, it makes sure that the diffusivity dynamics
is stationary, with a given correlation time. Second, the ensuing distribution
$p_D(D)$ has exponential tails, thus guaranteeing the emergence of the Laplace-like
distribution for $P(\mathbf{r},t)$ at short times, as we will show. At long times,
the above choice corresponds to a particle moving with an effective diffusivity
$\langle D\rangle$, and thus leads to the crossover to the long time Gaussian
behavior of $P(\mathbf{r},t)$. The above set of Langevin equations not only fulfill
these requirements but also allows for an analytical solution, as shown below.

For simplicity, we introduce dimensionless units via the transformations $t\to t/
\tau$ and $x\to x/(\sigma\tau)$ (and similar for $y$ and $z$). The process $\mathbf{
Y}(t)$ is renormalized according to $\mathbf{Y}\to\sigma\tau^{1/2}\mathbf{Y}$.
As detailed in Appendix \ref{app_dim} we then obtain the set of stochastic
equations
\begin{subequations}
\label{dimlesslang}
\begin{eqnarray}
\label{dlllang1}
\frac{d}{dt}\mathbf{r}(t)&=&\sqrt{2D(t)}\bm{\xi}(t),\\
\label{dlllang2}
D(t)&=&\mathbf{Y}^2(t),\\
\label{dlllang3}
\frac{d}{dt}\mathbf{Y}(t)&=&-\mathbf{Y}+\eta(t).
\end{eqnarray}
\end{subequations}
for our minimal diffusing diffusivity model.

We note that the above minimal model for the diffusing diffusivity allows different
choices for the number of components of $\mathbf{Y}(t)$.
The number $n$ is thus essentially
a free parameter of the model. It defines the number of `modes' necessary to describe
the random process $D(t)$. This is actually another advantage of the present approach,
since it provides additional flexibility.

In the Discussion section we will show that the above compound process is
analogous to the Heston model \cite{heston} and thus a special case of the
Cox-Ingersoll-Ross (CIR) model \cite{cir}, which are widely used for return
dynamics in financial mathematics. Our approach therefore has a wider appeal
beyond stochastic particle dynamics.

\subsection*{Properties of the Ornstein-Uhlenbeck process}

The stochastic equation (\ref{dlllang3}) contains a linear restoring term,
corresponding to the motion of the process $\mathbf{Y}$ in a centered harmonic
potential. The formal solution of this Ornstein-Uhlenbeck process reads
\begin{equation}
\mathbf{Y}(t)=\mathbf{Y}_0e^{-t}+\int_0^t\bm{\eta}(t')e^{-(t-t')}dt'.
\end{equation}
The associated autocorrelation function is
\begin{eqnarray}
\nonumber
\langle\mathbf{Y}(t_1)\mathbf{Y}(t_2)\rangle&=&\mathbf{Y}_0^2e^{-(t_1+t_2)}
+e^{-(t_1+t_2)}\\
\nonumber
&&\times\int_0^{t_1}
dt_1'\int_0^{t_2}dt_2'\langle\bm{\eta}(t_1')\bm{\eta}(t_2')\rangle e^{t_1'+t_2'}\\
&&\hspace*{-1.8cm}
=\mathbf{Y}_0^2e^{-(t_1+t_2)}+\frac{n}{2}\left(e^{-|t_2-t_1|}-e^{-(t_1+t_2)}\right).
\end{eqnarray}
Thus, for long times ($t_1+t_2\to\infty$), we find the exponential decay
\begin{equation}
\label{autocorr}
\langle\mathbf{Y}(t_1)\mathbf{Y}(t_2)\rangle\sim\frac{n}{2}e^{-|t_2-t_1|}
\end{equation}
of the autocorrelation, and thus the stationary variance
\begin{equation}
\label{meandiff}
\langle\mathbf{Y}^2(t)\rangle=\langle D\rangle_{\mathrm{st}}=\frac{n}{2}.
\end{equation}

We note that the Fokker-Planck equation for this Ornstein-Uhlenbeck process
reads
\begin{equation}
\frac{\partial}{\partial t}f(\mathbf{Y},t)=\frac{\partial}{\partial\mathbf{Y}}
\Big(\mathbf{Y}f(\mathbf{Y},t)\Big)+\frac{1}{2}\frac{\partial^2}{\partial\mathbf{
Y}^2}f(\mathbf{Y},t).
\end{equation}
The distribution $f(\mathbf{Y},t)$ converges to the normalized equilibrium
Boltzmann form
\begin{equation}
\label{fst}
f_{\mathrm{st}}(\mathbf{Y})=\frac{1}{\pi^{n/2}}e^{-\mathbf{Y}^2}.
\end{equation}

In what follows and in our simulations we assume that the initial condition
$\mathbf{Y}_0$ is taken randomly from the equilibrium distribution (\ref{fst}).
Then, the process $\mathbf{Y}(t)$ becomes stationary starting from $t=0$, and
Eq.~(\ref{autocorr}) is exact at all times.

The stationary diffusivity distribution $p_D(D)$ encoded in Eq.~(\ref{fst}) in
terms of the variable $\mathbf{Y}(t)$ can then be obtained as follows.

\textbf{(i)}
In dimension $n=1$, the variance of $Y$ in the stationary state is
$\langle Y^2\rangle_{\mathrm{st}}=1/2$
and the mapping to $p_D(D)$ reads
\begin{equation}
p_D^{\mathrm{st}}(D)=\int_{-\infty}^{\infty}f_{\mathrm{st}}(Y)\delta\left(D-Y^2
\right)dY=\frac{1}{\sqrt{\pi D}}e^{-D}.
\end{equation}
In dimensional form we have
\begin{equation}
\label{diffdiffexpo}
p_D^{\mathrm{st}}(D)=\frac{1}{\sqrt{\pi D_{\star}D}}e^{-D/D_{\star}},
\end{equation}
with
\begin{equation}
\label{dimd}
D_{\star}=\frac{x_0^2}{t_0}=\sigma^2\tau.
\end{equation}

From comparison with the direct superstatistical approach in Section \ref{superstat}
we see that the pure exponential form (\ref{expo0}) in our diffusing diffusivity
model is being modified by the additional prefactor $1/D^{1/2}$. From numerical
comparison, however, the exponential dependence is dominating and thus the result
(\ref{diffdiffexpo}) practically indistinguishable from (\ref{expo0}) for
sufficiently large $D$ values.

\textbf{(ii)} For $n=2$, the stationary state variance of $\mathbf{Y}$ is
$\langle\mathbf{Y}^2\rangle_{\mathrm{st}}=1$,
the mapping from $f_{\mathrm{st}}(\mathbf{Y})$ to
$p_D^{\mathrm{st}}(D)$ reads
\begin{equation}
p_D^{\mathrm{st}}(D)=2\pi\int_0^{\infty} Yf_{\mathrm{st}}(\mathbf{Y})\delta
\left(D-Y^2\right)dY=e^{-D},
\end{equation}
where $Y=|\mathbf{Y}|$ and $2\pi Yf_{\mathrm{st}}(\mathbf{Y})=2Y\exp(-Y^2)$.
Moreover, we made use of the property
\begin{equation}
\delta\left(Y^2-D\right)=\frac{1}{2\sqrt{D}}\Big[\delta\left(Y+\sqrt{D}\right)
+\delta\left(Y-\sqrt{D}\right)\Big].
\end{equation}
of the $\delta$-function. In dimensional units, we have
\begin{equation}
\label{2dpd}
p_D^{\mathrm{st}}(D)=\frac{1}{D_{\star}}e^{-D/D_{\star}},
\end{equation}
in conjunction with relation (\ref{dimd}).

\textbf{(iii)} Finally, for $n=3$ we have $\langle\mathbf{Y}^2\rangle_{\mathrm{st}}=
3/2$ and
\begin{equation}
p_D^{\mathrm{st}}(D)=4\pi\int_0^{\infty} Y^2f_{\mathrm{st}}(\mathbf{Y})\delta
\left(D-Y^2\right)=\frac{2\sqrt{D}}{\sqrt{\pi}}e^{-D},
\end{equation}
where $4\pi Y^2f_{\mathrm{st}}(\mathbf{Y})=4\pi^{-1/2}Y^2\exp(-Y^2)$. In
dimensional form,
\begin{equation}
\label{3dpd}
p_D^{\mathrm{st}}(D)=\frac{2\sqrt{D}}{\sqrt{\pi D_{\star}^3}}e^{-D/D_{\star}}.
\end{equation}

\section{Subordination concept for diffusing diffusivities}
\label{sec_subord}

Subordination, introduced by Bochner \cite{bochner}, is an important concept in
probability theory \cite{feller}. Simply
put, a subordinator associates a random time increment with the number of steps
of the subordinated stochastic process. For instance, continuous time random walks
with power-law distributions of waiting times can be described as a Brownian
motion in terms of the number of steps of the process, while the random waiting
times are introduced in terms of a L{\'e}vy stable subordinator, as originally
formulated by Fogedby \cite{fogedby} and developed as a stochastic representation
of the fractional Fokker-Planck equation \cite{report} and generalized
master equations for continuous time random walk models
\cite{igor,friedrich,friedricha,friedrichb}.

Here we apply and extend the subordination concept to a new class of random
diffusivity based stochastic processes. Our results for our minimal model of
diffusive diffusivities demonstrates that the subordination approach leads to a
superstatistical solution at times shorter than typical diffusivity correlation
times.

To start with, we note that the stochastic probability density function $\overline{
P}(\mathbf{r},t)=\overline{P}(\mathbf{r},t|D(t))$ fulfills the diffusion equation
\begin{equation}
\frac{\partial}{\partial t}\overline{P}(\mathbf{r},t)=D(t)\nabla^2\overline{P}
(\mathbf{r},t).
\end{equation}
With this in mind we can rewrite the Langevin equation (\ref{dlllang1}) in the
subordinated form
\begin{subequations}
\label{subo}
\begin{eqnarray}
\label{subo1}
\frac{d}{d\tau}\mathbf{r}(\tau)&=&\sqrt{2}\xi(\tau)\\
\label{subo2}
\frac{d}{dt}\tau(t)&=&D(t).
\end{eqnarray}
\end{subequations}
After this change of variables the Green function of the diffusion equation
has the form
\begin{equation}
\label{green}
G(\mathbf{r},\tau)=\frac{1}{\sqrt{(4\pi\tau)^d}}\exp\left(-\frac{r^2}{4
\tau}\right)
\end{equation}
with $r=|\mathbf{r}|$. The path variable $\tau$, for any given instant of time $t$,
according to Eq.~(\ref{subo2}) is a random quantity. In order to calculate the
probability density $P(\mathbf{r},t)$ of the variable $\mathbf{r}$ at time $t$ we
need to eliminate the path variable $\tau$. This is achieved by averaging the Green
function (\ref{green}) over the distribution of $\tau$ in the form
\begin{equation}
\label{subord}
P(\mathbf{r},t)=\int_0^{\infty}T_n(\tau,t)G(\mathbf{r},\tau)d\tau.
\end{equation}
Here $T_n(\tau,t)$ is the probability density function of the process
\begin{equation}
\label{tau_subord}
\tau(t)=\int_0^tD(t')dt'=\int_0^t\mathbf{Y}^2(t')dt'.
\end{equation}
Equation (\ref{subord}) is but the well known subordination formula, implying the
following: the probability for the walker to arrive at position $\mathbf{r}$ at
time $t$ equals the probability of being at $\tau$ on the path at time $t$,
multiplied by the probability of being at position $\mathbf{r}$ for this path
length $\tau$, summed over all path lengths \cite{fogedby}.

By help of relation (\ref{subord}) we write the Fourier transform
\begin{equation}
\hat{P}(\mathbf{k},t)=\int_{-\infty}^{\infty}e^{i\mathbf{k}\cdot\mathbf{r}}
P(\mathbf{r},t)d\mathbf{r}
\end{equation}
in the subordinated form
\begin{eqnarray}
\nonumber
\hat{P}(\mathbf{k},t)&=&\int_0^{\infty}T_n(\tau,t)\hat{G}(\mathbf{k},\tau)d\tau\\
&=&\int_0^{\infty}T_n(\tau,t)e^{-k^2\tau}d\tau=\tilde{T}_n(\mathbf{k}^2,t),
\label{fourierp}
\end{eqnarray}
with $k=|\mathbf{k}|$.
Thus, the Fourier transform of $P(\mathbf{r},t)$ is expressed in terms of the Laplace
transform $\tilde{T}_n$ of the density function $T_n(\tau,t)$ with respect to $\tau$,
\begin{equation}
\label{subord_square}
\tilde{T_n}(s,t)=\int_0^{\infty}e^{-s\tau}T_n(\tau,t)d\tau,
\end{equation}
with argument $s=k^2$.

\begin{figure}
\begin{center}
\includegraphics[width=8.8cm]{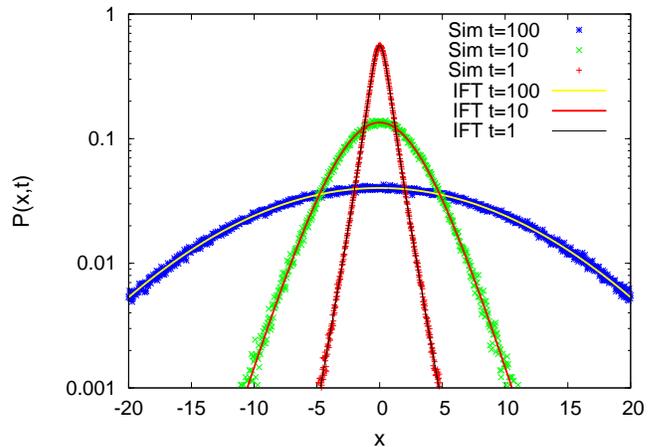}
\end{center}
\caption{Probability density function $P(x,t)$ in $d=n=1$ at longer times in
dimensionless form ($\sigma=\tau=D_{\star}=1$).
We compare results from simulations (Sim) of the set of Langevin equations
(\ref{dlllang1}) to (\ref{dlllang3}), represented by the symbols, with the direct
inverse Fourier transform (IFT) of result (\ref{pdfk}). Excellent agreement is
observed.}
\label{fig_pdflong}
\end{figure}

The subordination approach established here introduces a superior flexibility
into the diffusing diffusivity model. By specific choice of the subordinator
density $T_n(\tau,t)$ we may study a broad class of normal and anomalous
diffusion processes caused by diffusivities, that are randomly varying in
time and/or space. In turn, the advantage of our minimal model for diffusing
diffusivities introduced here is, that the process $\tau(t)$ is the integrated
square of the Ornstein-Uhlenbeck process, for which in the one-dimensional case
$n=1$ the Laplace transform of the probability density function is known
\cite{dankel},
\begin{eqnarray}
\nonumber
\tilde{T}_1(s,t)&=&\exp(t/2)\bigg/\Bigg[\frac{1}{2}\left(\sqrt{1+2s}+\frac{1}{\sqrt{
1+2s}}\right)\\
&&\hspace*{-0.8cm}
\times\sinh\left(t\sqrt{1+2s}\right)+\cosh\left(t\sqrt{1+2s}\right)\Bigg]^{1/2}.
\label{subordinator}
\end{eqnarray}
We thus directly obtain the exact analytical result for the Fourier transform
\begin{eqnarray}
\nonumber
\hat{P}(k,t)&=&\exp(t/2)\bigg/\Bigg[\frac{1}{2}\left(\sqrt{1+2k^2}+\frac{1}{\sqrt{
1+2k^2}}\right)\\
&&\hspace*{-1.2cm}
\times\sinh\left(t\sqrt{1+2k^2}\right)+\cosh\left(t\sqrt{1+2k^2}\right)
\Bigg]^{1/2}.
\label{pdfk}
\end{eqnarray}
of the probability density function $P(x,t)$. The inverse Fourier transform
can be performed numerically. Figure \ref{fig_pdflong} demonstrates excellent
agreement between this result and simulations of the stochastic starting
equations (\ref{dlllang1}) to (\ref{dlllang3}). Below we provide analytical estimates
of $P(x,t)$ for short and long times and establish a connection of the subordination
approach with the superstatistical framework.

Our approach can be easily generalized to the case of the $n$-dimensional
Ornstein-Uhlenbeck process considered by Jain and Sebastian \cite{klsebastian}.
Namely, let us consider
\begin{equation}
D(t)=\mathbf{Y}^2(t),
\end{equation}
where $\mathbf{Y(t)}=\{Y_1(t),\ldots,Y_n(t)\}$ is a $n$-dimensional
Ornstein-Uhlenbeck process. Since the components of $\mathbf{Y(t)}$ are
independent and
\begin{eqnarray}
\nonumber
\tau(t)&=&\int_0^t\mathbf{Y}^2(t)dt'\\
&=&\int_0^t\left(Y_1^2(t')+Y_2^2(t')+\ldots+Y_n^2(t')\right)dt',
\end{eqnarray}
the Laplace transform $\tilde{T}_n(s,t)$ and the characteristic function
$\hat{P}(\mathbf{k},t)$ are simply $n$-fold products of identical, one-dimensional
functions (\ref{subordinator}) and (\ref{pdfk}), respectively:
\begin{eqnarray}
\nonumber
\tilde{T}_n(s,t)&=&\exp(nt/2)\bigg/\Bigg[\frac{1}{2}\left(\sqrt{1+2s}+\frac{1}{\sqrt{
1+2s}}\right)\\
&&\hspace*{-0.8cm}
\times\sinh\left(t\sqrt{1+2s}\right)+\cosh\left(t\sqrt{1+2s}\right)\Bigg]^{n/2}.
\end{eqnarray}
We thus directly obtain the exact analytical result for the Fourier transform
\begin{eqnarray}\label{char func n dim}
\nonumber
\hat{P}(\mathbf{k},t)&=&\exp(nt/2)\bigg/\Bigg[\frac{1}{2}\left(\sqrt{1+2k^2}
+\frac{1}{\sqrt{1+2k^2}}\right)\\
&&\hspace*{-0.8cm}
\times\sinh\left(t\sqrt{1+2k^2}\right)+\cosh\left(t\sqrt{1+2k^2}\right)
\Bigg]^{n/2}.
\label{propfou}
\end{eqnarray}
This result is consistent with that of Eq.~(25) in
Ref.~\cite{klsebastian}, up to numerical factors, which appear due to the
difference in numerical coefficients entering the Ornstein-Uhlenbeck process.

\subsection{Brownian mean squared displacement and leptokurtic behavior}

The mean squared displacement and the fourth moment encoded in our minimal model
can be directly obtained from the Fourier transform (\ref{pdfk}) through
differentiation,
\begin{eqnarray}
\nonumber
\langle\mathbf{r}^2(t)\rangle&=&-\left.\nabla^2_{\mathbf{k}}\hat{P}(\mathbf{k},t)
\right|_{\mathbf{k}=0},\\
\langle\mathbf{r}^4(t)\rangle&=&\left.\nabla^4_{\mathbf{k}}\hat{P}(\mathbf{k},t)
\right|_{\mathbf{k}=0}.
\end{eqnarray}
For the isotropic case considered here the Laplace operator is defined as
\begin{equation}
\nabla^2_{\mathbf{k}}=\frac{1}{k^{d-1}}\frac{\partial}{\partial k}\left(k^{d-1}
\frac{\partial}{\partial k}\right).
\end{equation}
Expanding relation (\ref{fourierp}) for small $k$, we obtain up to the fourth order
\begin{eqnarray}
\nonumber
\hat{P}(\mathbf{k},t)&=&\int_0^{\infty}e^{-k^2\tau}T_n(\tau,t)d\tau\\
\nonumber
&=&1-k^2\int_0^{\infty}\tau T_n(\tau,t)d\tau\\
&&+\frac{k^4}{2}\int_0^{\infty}\tau^2T_n(\tau,t)d\tau+\ldots
\end{eqnarray}
From this we directly obtain the mean squared displacement
\begin{eqnarray}
\nonumber
\langle\mathbf{r}^2(t)\rangle&=&2d\int_0^{\infty}\tau T_n(\tau,t)d\tau=2d\langle
\tau\rangle\\
&=&-2d\left.\frac{\partial\tilde{T}_n(s,t)}{\partial s}\right|_{s=0}
\end{eqnarray}
and the fourth order moment
\begin{eqnarray}
\nonumber
\langle\mathbf{r}^4(t)\rangle&=&4d(2+d)\int_0^{\infty}\tau^2T_n(\tau,t)d\tau=4d
(2+d)\langle\tau^2\rangle\\
&=&4d(d+2)\left.\frac{\partial^2\tilde{T}_n(s,t)}{\partial s^2}\right|_{s=0}
\end{eqnarray}
With the results of Appendix \ref{appa} we find
\begin{equation}
\label{brown}
\langle\mathbf{r}^2(t)\rangle=dnt=2d\langle D\rangle_{\mathrm{st}}t,
\end{equation}
where $\langle D\rangle_{\mathrm{st}}$ is given by Eq.~(\ref{meandiff}), and
\begin{equation}
\label{fourth}
\langle\mathbf{r}^4(t)\rangle=4d(2+d)\langle D\rangle_{\mathrm{st}}\left[-\frac{
1-e^{-2t}}{2}+t+\langle D\rangle_{\mathrm{st}}t^2\right].
\end{equation}

\begin{figure}
\begin{center}
\includegraphics[width=8.8cm]{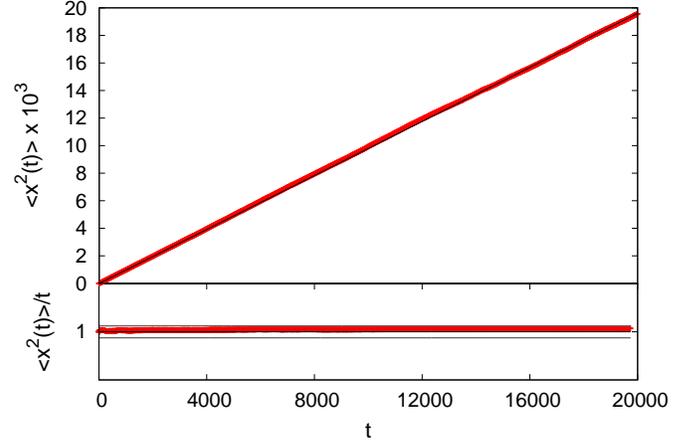}
\end{center}
\caption{Mean squared displacement $\langle x^2(t)\rangle$ obtained from
simulation of the set of Langevin equations (\ref{dlllang1}) to (\ref{dlllang3})
at $d=n=1$, corresponding to the red symbols (due to the density of points, these
rather appear as the thick red line), showing excellent agreement with the
Brownian law (\ref{brown}) shown by the (thin) full black line. In the bottom
panel we show $\langle x^2(t)\rangle/t$, demonstrating that the deviations
from the expected behavior are fairly. The grey lines (lower panel) show an interval
$[0.975,1.024]$ around unity, based on $10^6$ trajectories.}
\label{fig_msd}
\end{figure}

Eq.~(\ref{brown}) is the famed result of the normal Brownian, linear dispersion
of the mean squared displacement with time. In dimensional units the result
(\ref{brown}) reads
\begin{equation}
\label{browndim}
\langle\mathbf{r}^2(t)\rangle=dn\sigma^2\tau t=2d\langle D\rangle_{\mathrm{st}}
D_{\star}t.
\end{equation}
Fig.~\ref{fig_msd} demonstrates excellent agreement of the analytical result
(\ref{brown}) with direct simulations of the set of Langevin equations
(\ref{dlllang1}) to (\ref{dlllang3}) with respect to both slope and amplitude.

The deviation of the shape of a distribution function from a Gaussian can be
conveniently quantified in terms of the kurtosis
\begin{equation}
\label{kurt}
K=\frac{\langle\mathbf{r}^4(t)\rangle}{\langle\mathbf{r}^2(t)\rangle^2}.
\end{equation}
We note that the kurtosis is closely related to the (first) non-Gaussian parameter,
introduced in the classical text by Rahman \cite{rahman}. Inserting results
(\ref{brown}) and (\ref{fourth}) we obtain in the short time limit that
\begin{equation}
K\sim\left(1+\frac{2}{d}\right)\left(1+\frac{1}{\langle D\rangle_{\mathrm{st}}}
\right)=\left\{\begin{array}{ll}9,&d=1\\ 4,&d=2\\ 25/9,&d=3\end{array}\right.
\end{equation}
for the choice $d=n$. At long times,
\begin{equation}
K\sim\left(1+\frac{2}{d}\right)=\left\{\begin{array}{ll}3,&d=1\\
2,&d=2\\ 5/3,&d=3\end{array}\right..
\end{equation}
The first relation characterizes exponential distributions according to
Equations (\ref{short1}), (\ref{short2}), and (\ref{short3}) derived below,
whereas the second result coincides exactly with the kurtosis of the
multidimensional Gaussian distribution (note that this kurtosis does not depend on $n$). Taking along the next higher term in
the long time expansion, we observe that the leptokurtosis vanishes as $\simeq
1/t$ in the long time limit,
\begin{equation}
\label{leptosim}
K\sim1+\frac{2}{d}+\frac{2+d}{d\langle D\rangle_{\mathrm{st}}t}.
\end{equation}

\begin{figure}
\begin{center}
\includegraphics[width=8.8cm]{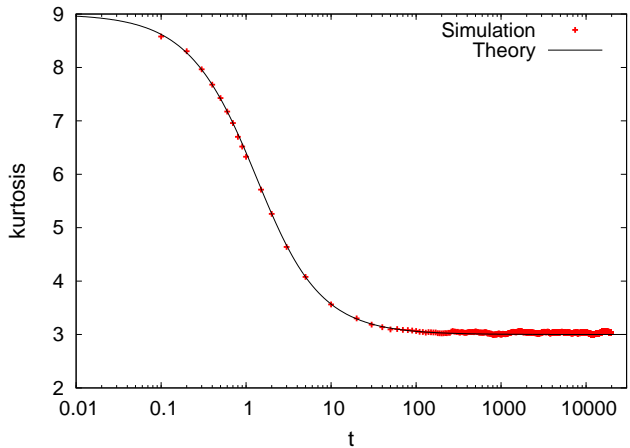}
\end{center}
\caption{The Kurtosis $K$ for $d=n=1$ defined in Equation (\ref{kurt}), based on
relations (\ref{brown}) and (\ref{fourth}) is shown by the full line, simulations
results are represented by the symbols. The crossover occurs at the correlation
time $\tau=1$. Over the entire displayed time range the agreement
between theory and simulations is excellent.}
\label{fig_kurt}
\end{figure}

In Fig.~\ref{fig_kurt} we compare the analytical result (\ref{kurt}) for the kurtosis
for $d=1$ based on Equations (\ref{brown}) and (\ref{fourth}) with simulations,
showing excellent agreement from the short time behavior all the way to the
saturation plateau at the Gaussian value $K=3$. We note that the crossover
time from strongly leptokurtic to Gaussian behavior occurs at $t\approx 1$, which
in dimensional units corresponds to the correlation time of the diffusing
diffusivity process.
Thus in experiments the behavior of the kurtosis as function of time provides
a direct means to extract the correlation time of $D(t)$, which also corresponds
to the crossover time from the exponential to the Gaussian behavior of the
probability density $P(x,t)$, as will be shown below. We also note that when the
process $\mathbf{Y}(t)$ is very highly dimensional and thus $\langle D\rangle_{
\mathrm{st}}$ large, the exponential tails of $P(\mathbf{r},t)$ do not exist, see
Appendix \ref{appb}.

We now derive explicit analytical results for the probability density function
$P(\mathbf{r},t)$ in the short and long time limits, starting with the short time
limit and its relation to the superstatistical formulation of the diffusing
diffusivity. In our further exemplary calculations we take $n=d$. However, in
Appendix C we discuss the situations when $n=1$ and $d=2$, as well as when $n$
goes to infinity while $d$ stays finite.

\subsection{Short time limit}

First we concentrate on the shape of the density $P(\mathbf{r},t)$ 
in the short time limit $t\ll\tau$.
In dimensionless units this means that we consider the asymptotic behavior of
the characteristic function (\ref{propfou}) under the condition $t\ll1$,
for which 
\begin{equation}
\sinh\left(t\sqrt{1+2k^2}\right)\sim t\sqrt{1+2k^2},\quad \cosh\left(t\sqrt{1+2k^2}
\right)\sim1,
\end{equation}
Together with expression (\ref{propfou}) we thus find
\begin{equation}
\label{fouprop}
\hat{P}(\mathbf{k},t)\sim\frac{(1+t)^{n/2}}{(1+[1+k^2]t)^{n/2}}\sim t^{-n/2}
\left(k^2+\frac{1}{t}\right)^{-n/2}.
\end{equation}
This expression is indeed normalized, $\hat{P}(\mathbf{k}=0,t)=1$. We can thus
perform the inverse Fourier transform to obtain $P(\mathbf{r},t)$ in the 
short time limit.

\textbf{(i)} For one dimension $d=n=1$ we find
\begin{equation}
\label{pshort}
P(x,t)\sim\frac{1}{\pi t^{1/2}}\int_0^{\infty}\frac{\cos(kx)}{\left(k^2+1/t\right)
^{1/2}}dk=\frac{1}{\pi t^{1/2}}K_0\left(\frac{x}{t^{1/2}}\right),
\end{equation}
in terms of the Bessel function \cite{gradshteyn}
\begin{equation}
K_0(a\beta)=\int_0^{\infty}\frac{\cos(ax)}{\sqrt{x^2+\beta^2}}dx.
\end{equation}
Apart from the normalization we observe that from Eq.~(\ref{pshort}) we also
derive the Brownian behavior $\langle x^2(t)\rangle=t=2\langle D\rangle_{
\mathrm{st}}t$, in accordance with Eqs.~(\ref{meandiff}) and (\ref{brown}).

Keeping in mind that here we are pursuing the large value limit of the scaling
variable $z=xt^{-1/2}\gg 1$, we expand the Bessel function in the form
\cite{abramowitz}
\begin{equation}
K_0(z)\sim\sqrt{\frac{\pi}{2z}}e^{-z}.
\end{equation}
We thus find the asymptotic result
\begin{equation}
\label{short1}
P(x,t)\sim\frac{1}{\sqrt{2\pi|x|t^{1/2}}}\exp\left(-\frac{|x|}{t^{1/2}}\right).
\end{equation}
This expression reproduces the exponential shape of the probability density
function $P(x,t)$ of the diffusing diffusivity model, with the power-law
correction $|x|^{-1/2}$.

\begin{figure}
\begin{center}
\includegraphics[width=8.8cm]{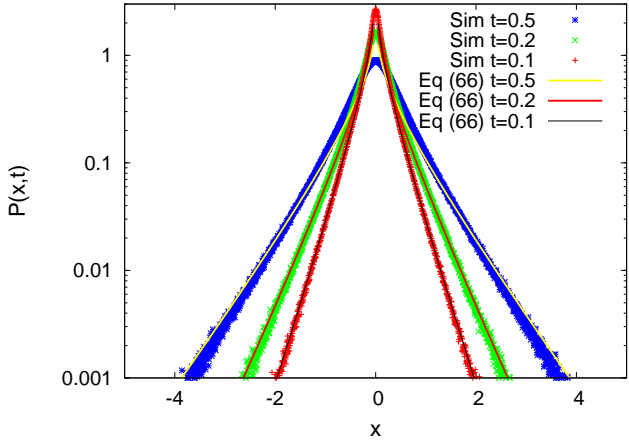}
\end{center}
\caption{Probability density function $P(x,t)$ for $d=n=1$ at short times in
dimensionless form ($\sigma=\tau=D_{\star}=1$). We compare
results from simulations of the set of Langevin equations (\ref{dlllang1}) to
(\ref{dlllang3}), represented by the symbols, with the explicit short time solution
(\ref{pshort}). Excellent agreement is observed, only for the longest time
$t=0.5$ the wings start to show some deviations.}
\label{fig_pdfshort}
\end{figure}

Figure \ref{fig_pdfshort} demonstrates excellent agreement of our short time
result (\ref{pshort}) and simulations. For the longest simulated time the
wings of the distribution start to show some deviations, indicating that in
this case the short time limit is no longer fully justified.

\textbf{(ii)} For $d=n=2$ we obtain
\begin{eqnarray}
\nonumber
P(\mathbf{r},t)&=&\int e^{i\mathbf{k}\cdot\mathbf{r}}\hat{P}(\mathbf{k},t)
\frac{d\mathbf{k}}{(2\pi)^2}\\
&=&\frac{1}{2\pi}\int_0^{\infty}kJ_0(kr)\hat{P}(k,t)dk,
\end{eqnarray}
and thus
\begin{equation}
\label{2dshort}
P(\mathbf{r},t)=\frac{1}{2\pi t}K_0\left(\frac{r}{\sqrt{t}}\right).
\end{equation}
Here we used the relation
\begin{equation}
\int_0^{\pi}\cos(kr\cos\varphi)d\varphi=\pi J_0(kr)
\end{equation}
in terms of the modified Bessel function $J_0$. The distribution is normalized
and encodes the Brownian behavior $\langle\mathbf{r}^2(t)\rangle=4t=4\langle
D\rangle_{\mathrm{st}}t$. Expanding the Bessel function $K_0$ as above we
find that
\begin{equation}
\label{short2}
P(\mathbf{r},t)\sim\frac{1}{2\sqrt{2\pi rt^{3/2}}}e^{-r/\sqrt{t}}.
\end{equation}

\textbf{(iii)} Finally, in $d=n=3$ the asymptotic probability density becomes
\begin{eqnarray}
\nonumber
P(\mathbf{r},t)&=&\int e^{i\mathbf{k}\cdot\mathbf{r}}\hat{P}(\mathbf{k},t)\frac{
d\mathbf{k}}{(2\pi)^3}\\
\nonumber
&=&\frac{1}{8\pi^3}\int_0^{\infty}k^2dk\int_0^{\pi}\sin
\theta d\theta\int_0^{2\pi}d\varphi e^{ikr\cos\theta}\\
\nonumber
&&\times t^{-3/2}\left(k^2+\frac{1}{t}\right)^{-3/2}\\
\nonumber
&=&\frac{1}{4\pi^2t^{3/2}}\int_0^{\infty}k^2dk\int_{-1}^1d\xi e^{ikr\xi}\left(
k^2+\frac{1}{t}\right)^{-3/2}\\
&=&\frac{1}{2\pi^2t^{3/2}}K_0\left(\frac{r}{\sqrt{t}}\right),
\label{3dshort}
\end{eqnarray}
and we have $\langle\mathbf{r}^2(t)\rangle=9t=6\langle D\rangle_{\mathrm{st}}t$.
Expansion of the Bessel function produces the asymptotic exponential behavior
\begin{equation}
\label{short3}
P(\mathbf{r},t)\sim\frac{1}{(2\pi)^{3/2}r^{1/2}t^{5/4}}e^{-r/\sqrt{t}}.
\end{equation}

As we will show in the following Subsection, in all dimensions the short time limit reproduces the superstatistical behavior. The subordination formulation,
the explicit result (\ref{pshort}) in terms of the Bessel function, and the
asymptotic exponential (Laplace) form (\ref{short1}) (and their multidimensional
analogs (\ref{2dshort}) and (\ref{short2}) to (\ref{short3})) constitute our first
main result. We note that the asymptotic forms (\ref{short1}), (\ref{short2}),
and (\ref{short3}) of the density function $P(x,t)$ by itself leads to the Brownian
scaling (\ref{brown}) of the corresponding mean squared displacement. When $n\neq d$
the exponential shape of the short time behavior is conserved while the subdominant
prefactors change, as shown for the case $d=2$ and $n=1$ in Appendix \ref{appb}.

\subsection{Relation to the superstatistical approximation}
\label{secsuperstat}

Above we formulated the concept of diffusing diffusivities in terms of coupled
stochastic equations for the particle position $\mathbf{r}(t)$ and the random
diffusivity $D(t)$. An alternative approach suggested in Refs.~\cite{granick,gary}
is that of the superstatistical distribution of the diffusivity, as laid out in
Section \ref{superstat}. In this superstatistical sense the overall distribution
function is given as the weighted average of a single Gaussian over the stationary
diffusivity distribution,
\begin{equation}
\label{super}
P_s(\mathbf{r},t)=\int_0^{\infty}p_D^{\mathrm{st}}(D)G(\mathbf{r},t|D)dD.
\end{equation}

\textbf{(i)} In dimension $d=n=1$ our minimal model produces with
Eq.~(\ref{diffdiffexpo})
\begin{eqnarray}
\nonumber
P_s(x,t)&=&\frac{1}{2\pi\sqrt{D_{\star}t}}\int_0^{\infty}\frac{1}{D}
\exp\left(-\frac{D}{D_{\star}}-\frac{x^2}{4Dt}\right)dD\\
&=&\frac{1}{\pi\sqrt{D_{\star}t}}K_0\left(\frac{|x|}{\sqrt{D_{\star}t}}\right).
\end{eqnarray}

\textbf{(ii)} In $d=n=2$, we have with Eq.~(\ref{2dpd})
\begin{eqnarray}
\nonumber
P_s(\mathbf{r},t)&=&\frac{1}{4\pi D_{\star}t}\int_0^{\infty}\frac{1}{D}\exp
\left(-\frac{D}{D_{\star}}-\frac{r^2}{4Dt}\right)dD\\
&=&\frac{1}{2\pi D_{\star}t}K_0\left(\frac{r}{\sqrt{D_{\star}t}}
\right).
\end{eqnarray}

\textbf{(iii)} Finally, in $d=n=3$, we find with Eq.~(\ref{3dpd})
\begin{eqnarray}
\nonumber
P_s(\mathbf{r},t)&=&\frac{2}{\pi^2(4D_{\star}t)^{3/2}}\int_0^{
\infty}\frac{1}{D}\exp\left(-\frac{D}{D_{\star}}-\frac{r^2}{4Dt}\right)dD\\
&=&\frac{1}{2\pi^2(D_{\star}t)^{3/2}}K_0\left(\frac{r}{\sqrt{D_{\star}t}}\right).
\end{eqnarray}

For all $d$ the mean squared displacement acquires the linear Brownian scaling in
time $\langle\mathbf{r}^2\rangle=2d \langle D\rangle_{\mathrm{st}}t$, as it should.

This is but exactly the result of our subordination scheme in the short time limit,
expressions (\ref{short1}), (\ref{2dshort}), and (\ref{3dshort}), written in
dimensional form. Thus, in our approach to the diffusing diffusivity the short
time regime of the subordination formalism leads directly to the superstatistical
result. The reason is as follows: At times less than the diffusivity correlation
time $\tau$ the diffusion coefficient does not change considerably, and the
subordination scheme describes an ensemble of particles, each diffusing with its own
diffusion coefficient. This mimics a spatially inhomogeneous situation, when the
local diffusion coefficient is random, but stays constant within confined spatial
domains. In this case the ensemble of particles moving in different domains exhibits
a superstatistical behavior, as assumed in the original works \cite{beck}. However,
in any system with finite patch sizes, we would not expect the particles to stay in
their local patch of diffusivity $D$ forever, thus violating the assumption of the
superstatistical approach. Our annealed approach in some sense delivers a mean field
approximation to the spatially disordered situation, and adequately describes the
transition from short time superstatistical behavior to the Gaussian probability
law at long times, which will be shown in the subsequent section. The full
consistency in the short time limit between the subordination approach and
superstatistics is our second main result.

\subsection{Long time limit}

We now turn to the long time limit encoded in the Fourier transform (\ref{pdfk}) of
the probability density $P(\mathbf{r},t)$, that is, the times larger than the
diffusivity correlation time $\tau$. In dimensionless units it corresponds to $t
\gg1$, and the hyperbolic functions assume the limiting behaviors
\begin{eqnarray}
\nonumber
\sinh\left(t\sqrt{1+2k^2}\right)&\sim&\cosh\left(t\sqrt{1+2k^2}\right)\\
&\sim&\frac{1}{2}\exp\left(t\sqrt{1+2k^2}\right).
\end{eqnarray}
Combined with result (\ref{char func n dim}) we find
\begin{equation}
\label{longtime}
\hat{P}(\mathbf{k},t)\sim\frac{2^{n/2}\exp\left(\frac{nt}{2}\left[1-\sqrt{1+2k^2}
\right]\right)}{\left(1+\frac{1}{2}\left[\sqrt{1+2k^2}+\frac{1}{\sqrt{1+2k^2}}
\right]\right)^{n/2}}.
\end{equation}
As in the short time limit above, this expression is normalized, $\hat{P}
(\mathbf{k}=0,t)=1$.

Now let us focus on the tails of the probability density $P(\mathbf{r},t)$,
corresponding to the limit $k\ll1$, for which Eq.~(\ref{longtime}) gives $\hat{P}
(\mathbf{k},t)\sim\exp(-nk^2t/2)=\exp\left(-\langle D \rangle_{\mathrm{st}}k^2t
\right)$, and thus
\begin{equation}\label{gaussian}
P(\mathbf{r},t)\sim\frac{1}{(4\pi\langle D\rangle_{\mathrm{st}}t)^{n/2}}\exp\left(
-\frac{\mathbf{r}^2}{4\langle D\rangle_{\mathrm{st}}t}\right).
\end{equation}
At long times the probability density function $P(\mathbf{r},t)$ assumes a Gaussian
form, with the effective diffusivity $\langle D\rangle_{\mathrm{st}}=n/2$. This is a
consequent result given the Ornstein-Uhlenbeck variation of the diffusivity
encoded in the starting equations (\ref{lang2}) and (\ref{lang3}): at
sufficiently long times the process samples the full diffusivity space and
behaves like an effective Gaussian process with renormalized diffusivity. The
explicit derivation of the crossover to the Gaussian behavior is our third main
result.

\begin{figure}
\includegraphics[width=8.8cm]{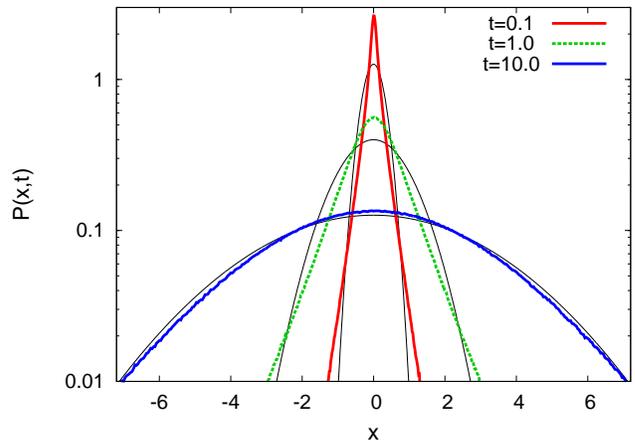}
\caption{Probability density function $P(x,t)$ for $d=n=1$ from simulations of the
Langevin equations (\ref{dimlesslang}) for three different times, in
dimensionless form ($\sigma=\tau=D_{\star}=1$). Comparison
with Gaussian distribution (\ref{gaussian}) demonstrates the strongly non-Gaussian
behavior at short times and the almost fully Gaussian shape at longer times.}
\label{bothlimits}
\end{figure}

Fig.~\ref{bothlimits} shows the crossover from the initial exponential to the
long time Gaussian behavior of the probability density function $P(x,t)$ for
$d=n=1$ by comparison to the Gaussian distribution (\ref{gaussian}) for short
time $t=0.1$, the crossover time $t=1.0$ and the longer time $t=10.0$.

\section{Bivariate Fokker-Planck equation and relation to the subordination
approach}

In this section we derive the Fokker-Planck equation corresponding to the set of
stochastic equations (\ref{dlllang1}) to (\ref{dlllang3}) of our diffusing diffusivity
model. We will also establish the relation to the subordination approach of
section \ref{sec_subord}. Note that we here restrict the discussion to the case
$d=n=1$, as higher dimensional cases are completely equivalent.

Following our notation we thus seek the Fokker-Planck equation for the bivariate
probability density function $f(x,y,t)$, which has the structure
\begin{equation}
\label{fpe}
\frac{\partial}{\partial t}f(x,y,t)=\mathscr{L}_yf(x,y,t)+y^2\frac{\partial^2}{
\partial x^2}f(x,y,t).
\end{equation}
The Fokker-Planck operator in $y$ reads
\begin{equation}
\mathscr{L}_y=\frac{\partial}{\partial y}y+\frac{1}{2}\frac{\partial^2}{\partial
y^2},
\end{equation}
and the marginal probability density function for $x$ is then
\begin{equation}
\label{marginal}
P(x,t)=\int_{-\infty}^{\infty}f(x,y,t)dy.
\end{equation}

To proceed further we introduce the joint probability density function $q(\tau,y,t)$
of the Ornstein-Uhlenbeck process $y(t)$ and its integrated square $\tau(t)$, given
by equation (\ref{tau_subord}). The corresponding system of stochastic equations
has the form
\begin{subequations}
\begin{eqnarray}
\frac{dy}{dt}&=&-y+\eta(t),\\
\frac{d\tau}{dt}&=&y^2,
\end{eqnarray}
\end{subequations}
and thus the bivariate Fokker-Planck equation governing the probability density
$q(\tau,y,t)$ reads
\begin{equation}
\label{fpe1}
\frac{\partial}{\partial t}q(\tau,y,t)=\mathscr{L}_yq(\tau,y,t)-y^2\frac{\partial}{
\partial\tau}q(\tau,y,t).
\end{equation}
Now we introduce an ansatz for the solution of equation (\ref{fpe}) of the form
\begin{equation}
\label{ansatz}
f(x,y,t)=\int_0^{\infty}G(x,\tau)q(\tau,y,t)d\tau,
\end{equation}
where $G(x,\tau)$ is the Gaussian probability density given by equation
(\ref{green}). Then, in accordance with equation (\ref{marginal}) the marginal
probability density $P(x,t)$ can be written in the subordination form of equation
(\ref{subord}), where
\begin{equation}
T(\tau,t)=\int_{\infty}^{\infty}q(\tau,y,t)dy
\end{equation}
is the marginal probability density function of the integrated square of the
Ornstein-Uhlenbeck process whose Laplace transform is given by equation
(\ref{subord_square}).

We now prove that the solution of equation (\ref{fpe}) can be presented in the
form (\ref{ansatz}). To that end we differentiate equation (\ref{ansatz}) and use
relation (\ref{fpe1}) to get
\begin{eqnarray}
\nonumber
\frac{\partial}{\partial t}f(x,y,t)&=&\int_0^{\infty}G(x,\tau)\frac{\partial}{
\partial t}q(\tau,y,t)d\tau\\
\nonumber
&&\hspace*{-1.2cm}
=\int_0^{\infty}G(x,\tau)\left[\mathscr{L}_yq(x,y,t)-y^2\frac{\partial}{\partial
\tau}q(x,y,t)\right]d\tau\\
\nonumber
&&\hspace*{-1.2cm}
=\mathscr{L}_y\left[\int_0^{\infty}G(x,\tau)q(\tau,y,t)d\tau\right]\\
&&\hspace*{-1.2cm}-y^2\int_0^{\infty}
G(x,\tau)\frac{\partial}{\partial\tau}q(x,y,t)d\tau.
\end{eqnarray}
We now apply relation (\ref{ansatz}) to the first term and integrate the second
term by parts, obtaining
\begin{eqnarray}
\nonumber
\frac{\partial}{\partial t}f(x,y,t)&=&\mathscr{L}_yf(x,y,t)-y^2\left\{q(\tau,y,t)
G(x,\tau)\Big|^{\tau=\infty}_{\tau=0}\right.\\
&&-\left.\int_0^{\infty}q(\tau,y,t)\frac{\partial}{\partial\tau}G(x,\tau)d\tau
\right\}.
\end{eqnarray}
Finally, with the relation $\partial G(x,\tau)/\partial\tau=\partial^2G(x,\tau)/
\partial x^2$, we see that
\begin{eqnarray}
\nonumber
\frac{\partial}{\partial t}f(x,y,t)&=&\mathscr{L}_yf(x,y,t)+y^2\frac{\partial^2}{
\partial x^2}f(x,y,t)\\
&&+y^2q(\tau=0,y,t)\delta(x).
\end{eqnarray}
The last term vanishes, as $\tau$ is the integrated square of $y(t)$, and thus we
arrive at equation (\ref{fpe}). Therefore we showed that the solution of the
bivariate Fokker-Planck equation (\ref{fpe}) can be presented in the form
(\ref{ansatz}), and consequently the marginal probability density function $P(x,t)$
can be written in the subordination form (\ref{subord}).

The connection of the bivariate Fokker-Planck equation (\ref{fpe}) for the Langevin
system (\ref{lang1}) to (\ref{lang3}) with the subordination approach represented
by relations (\ref{subo}) and (\ref{subord}) is our fourth main result.

\section{Discussion}

An increasing number of systems are reported in which the mean squared displacement
is linear in time, suggesting normal (Fickian) diffusion of the observed tracer
particles. Concurrently the (displacement) probability density function is
pronouncedly non-Gaussian. Normal diffusion with a Laplace distribution of particle
displacements was previously explained in a superstatistical approach by Granick and
coworkers \cite{granick}. The experimentally observed crossover to Gaussian
statistics at longer times was interpreted as a consequence of the central limit
theorem, kicking in at times longer then the correlation time of the diffusion
fluctuations \cite{granick}. Chubinsky and Slater \cite{gary} introduced the
diffusing diffusivity model and studied it numerically, concluding the crossover
from the initial exponential shape to a Gaussian with effective diffusivity. Jain
and Sebastian go further with the double Langevin approach \cite{klsebastian}. Here
we introduced a consistent minimal model for a diffusing diffusivity. We explicitly
obtain the Fourier transform of the full probability density function, from which we
derive the analytical short and long time limits. This allows us to determine the
dynamical crossover to the long time Gaussian shape of the probability density at
the correlation time of the fluctuating diffusivity. Moreover we demonstrate a full
consistency of our minimal model with the superstatistical approach, as well as with
the results of Jain and Sebastian. At the same time our model is more general and
flexible: phrasing the diffusing diffusivity approach in terms of a subordination
concept we endow our model with an extremely flexible basis, such that a wide range
of different statistics for the diffusivity can be included.

We also obtained the bivariate Fokker-Planck equation for this diffusing diffusivity
process and expressed its solution in terms of the subordination integral.
Excellent agreement with simulations is provided for the probability
density function, the Brownian scaling of the mean squared displacement, and the
kurtosis of the probability density function.

We are confident that this subordination integral formulation of the diffusing
diffusivity model will prove useful for experimentalists observing such dynamics.
The model can be calibrated with respect to the two parameters $\tau$ and
$\sigma$, which can be obtained from experiment by analyzing the time-dependence
of the mean squared displacement and the kurtosis. The latter provides information
on the typical diffusivity correlation time $\tau$ (see Fig.~\ref{fig_kurt}),
whereas the former allows one to estimate the parameter $\sigma$, see
Eq.~(\ref{browndim}). Moreover, measurement of the diffusivity distribution, as
can be directly obtained experimentally \cite{spako}, provides the value of the
parameter $D_{\star}=\sigma^2\tau$ according to Eq.~\ref{diffdiffexpo}.
Additionally, the possibility to include a different
dimensionality $n$ for the subordinating process $\mathbf{Y}(t)$ allows
for fine-tuning of the model to match the experimentally observed probability
density function, which might be necessary when the model is used for quantitative
predictions. As we show in the example in Appendix \ref{appb}1 the difference in
the dimensionality $n$ of the process $\mathbf{Y}(t)$ does not change the
dominant exponential behavior at short times but affects the prefactors.
Similarly the prefactors of the exponential in the diffusivity distribution
$p_D^{\mathrm{st}}(D)$ is affected by the concrete value of $n$.
This fact may be employed to account for the deviations from the pure exponential
shape of the probability distribution also reported in \cite{granick}.

An intriguing question emerging from our analysis concerns the physical origin
of the dimensionality $n$ of the subordinating process $\mathbf{Y}(t)$. Intuitively,
one might argue that the diffusivity and thus $\mathbf{Y}(t)$ should have as many
components as spatial directions in the particle trajectory $\mathbf{r}(t)$, i.e.,
$d=n$. This is the case considered in the derivations in Section \ref{sec_subord}.
However, the mode concept for $D(t)$ introduced here may also have a more fundamental
physical meaning. As we showed here, the value $n$ affects the details of the shape
of $P(\mathbf{r},t)$ as well as $p_D^{\mathrm{st}}(D)$, and for large $n$ values the
short time exponential shapes may even be fully suppressed. Advanced experiments
allowing one to determine $n$ from the exact shape of the probability densities
$P(\mathbf{r},t)$ and $p_D^{\mathrm{st}}(D)$ will provide important clues
concerning this question.

Possible generalizations may include diffusing diffusivity models with additional
deterministic time dependence of the diffusivity \cite{sbmdiff1} or non-Gaussian
anomalous viscoelastic diffusion in crowded membranes \cite{membranesa}, which
contrasts Gaussian anomalous viscoelastic diffusion
in non-crowded membranes \cite{membranesc}. Of course, the diffusing diffusivity
concept is a first step in capturing the full spatiotemporal disorder of complex
systems. Ultimately, a full description of spatial and temporal stochasticity in
terms of a random diffusivity $D(x,t)$ will be desired.

Let us put the diffusing diffusivity approach into context with other popular
models with distributed transport coefficients. Typically these are constructed to
describe anomalous diffusion processes. Another model is scaled Brownian
motion, in which the diffusivity is a deterministic, power-law function of time
\cite{lim,fulinski}. On a stochastic level scaled Brownian motion appears naturally
in granular gases \cite{gasbook}, in which non-ideal collisions effect a decrease of
the system's temperature (kinetic energy). Scaled Brownian motion is non-ergodic
and displays a massively delayed overdamping transition \cite{sbm}. Heterogeneous
diffusion processes employ a continuous, deterministic space dependence of the
diffusivity and lead to non-ergodic and ageing dynamics \cite{hdp,hdpa,fulinski}. In contrast to
these models random diffusivity approaches also have a considerable history. Thus
segregation in solids in the context of radiation was described by such an approach 
\cite{dubinko}, and Brownian motion in media with fluctuating friction coefficient,
temperature fluctuations, or randomly interrupted diffusion were used to describe,
for instance, randomly stratified media \cite{luczka-talkner}. A random diffusivity
approach was elaborated to consider light scattering in a continuous medium with
fluctuating dielectric constant \cite{kravtsov}. Motivated by the comparison of
diffusion processes assessed by different modern measurement techniques, the concept
of \emph{microscopic single-particle diffusivity\/} was developed \cite{radons}.
In \cite{lapeyre} the diffusivity varies randomly but is constant on patches of
random sizes. Such random patch model show non-ergodic subdiffusion due to the
diffusivity effectively changing at random times with a heavy-tailed distribution.
Intermittency between two values of the diffusivity were also considered
\cite{sbmdiff}.
Finally, we mention that a deterministic time dependence of the diffusivity was
combined with a random diffusivity in \cite{sbmdiff1}. As seen in several experimental
studies already, to describe stochastic particle motion in real complex systems such
as living biological cells, combinations of different stochastic mechanisms are
necessary to capture the observed dynamics \cite{goldinga}. Thus also the
diffusing diffusivity picture may need to be complemented by other processes, as
we saw in the example of non-Gaussian viscoelastic subdiffusion based on the
observations in \cite{spako,membranesa}.

Despite the wealth of established stochastic processes the diffusing diffusivity
model has quite unique properties. Thus the crossover from a short time exponential
shape to Gaussian statistics at longer times, while the MSD remains linear and thus
classifies normal (Fickian) diffusion, cannot be captured by existing models. Of
course, crossovers between non-Gaussian to Gaussian probability density functions
may be grasped by truncated continuous time random walks or distributed order
fractional diffusion equations \cite{truncate}. However, in these models also the
MSD exhibits a crossover from anomalous to normal diffusion. In this sense we
believe that the diffusing diffusivity model and its potential generalizations on
the basis of our subordination approach will emerge as a new paradigm in the theory
of stochastic processes.

We conclude with pointing out that the diffusing diffusivity model developed here
is closely related to the Cox-Ingersoll-Ross (CIR) model for monetary returns which
is widely used in financial mathematics \cite{cir}. To show this relation let us
write the Langevin equation (\ref{lang3}) for the Ornstein-Uhlenbeck process as
\begin{equation}
dY_i=-\frac{1}{\tau}Y_idt+\sigma dW_i(t),
\end{equation}
where $i=1,\ldots,n$ and $W_i(t)$ is the Wiener process with variance $1/2$. Our
aim is to design a Langevin equation in the It\^{o} form for the squared
Ornstein-Uhlenbeck process in $n$ dimensions,
\begin{equation}
D(t)=\sum_{i=1}^nY_i^2(t).
\end{equation}
To this end we employ the It\^{o} formula of differentiation to the function of
a $n$-dimensional vector \cite{gardiner} to find
\begin{equation}
dD=\frac{2}{\tau}\left(\frac{n\sigma^2\tau}{2}-D\right)+2\sigma\sqrt{D}dW(t).
\end{equation}
This is but the stochastic differential equation of the CIR process describing
the time evolution of interest rates \cite{cir}. The same process is used in the
Heston model specifying the evolution of stochastic volatility of a given asset
\cite{heston}. Our results for the subordination approach should therefore also
be relevant to financial market modeling. Indeed, the technique of subordination,
which is closely related to random time changes, is a very common concept in
financial mathematics \cite{clark}.

\appendix

\section{Superstatistics with modified exponential diffusivity distribution}
\label{appc}

Consider the Gaussian probability density function typical for viscoelastic
subdiffusion in the overdamped limit,
\begin{equation}
G_{\alpha}(x,t|D_{\alpha})=\frac{1}{\sqrt{4\pi D_{\alpha}t^{\alpha}}}\exp\left(
-\frac{x^2}{4D_{\alpha}t^{\alpha}}\right),
\end{equation}
which is equivalent to fractional Brownian motion \cite{pccp}. The associated
mean squared displacement is $\langle x^2(t)\rangle=2D_{\alpha}t^{\alpha}$. For
the superstatistical distribution of the generalized diffusion coefficient we
choose the modified exponential
\begin{equation}
p_D(D_{\alpha})=\frac{1}{\Gamma(1+1/\kappa)D_{\alpha}^{\star}}
\exp\left(-\left[\frac{D_{\alpha}}{D_{\alpha}^{\star}}\right]^{\kappa}
\right).
\end{equation}

\begin{figure}
\includegraphics[width=8cm]{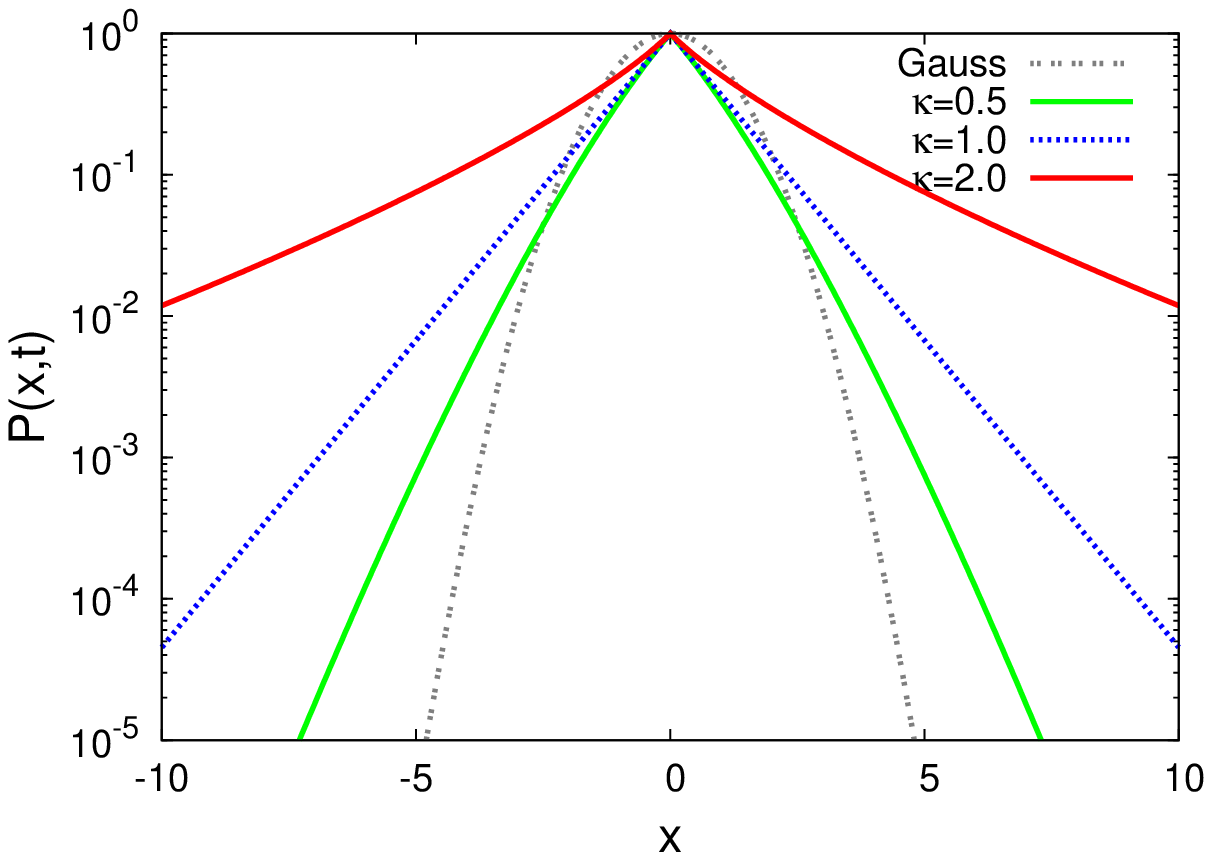}
\includegraphics[width=8cm]{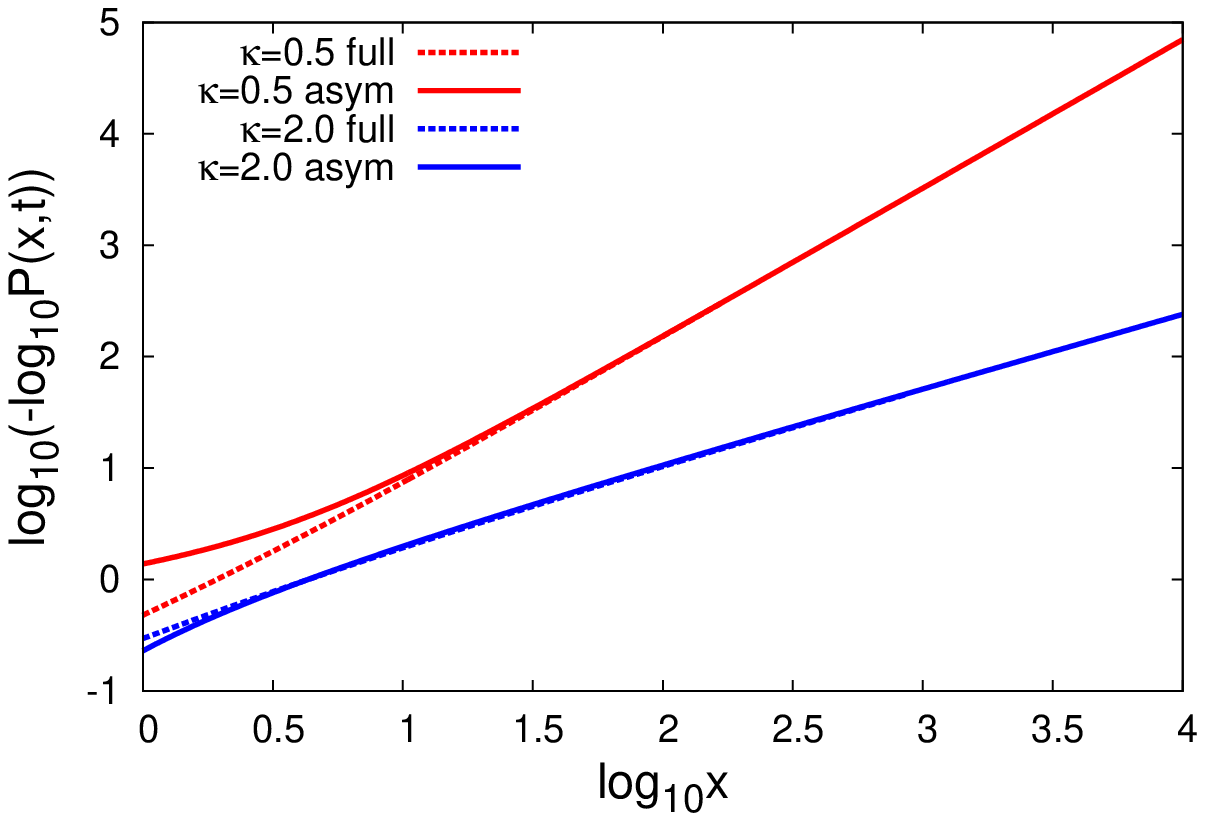}
\caption{Top: Superstatistical probability density function $P_s(x,t)$ according
to Equation (\ref{stretched_conv}) from numerical integration, for exponents
$\kappa=0.5$, $1$, and $2$ (see the Figure key). Bottom: convergence of the full
numerical solution to the analytical asymptotic form (\ref{stretched_as}) for
$\kappa=0.5$ and $2$. All distributions are drawn for $t=1$ (a.u.).}
\label{fig_stretched}
\end{figure}

The resulting probability density function
\begin{equation}
\label{stretched_conv}
P_s(x,t)=\int_0^{\infty}p_D(D_{\alpha})G_{\alpha}(x,t|D_{\alpha})dD_{\alpha}
\end{equation}
with $D_{\alpha}/D_{\alpha}^{\star}\to\tilde{D}$ and $\lambda=x^2/[4D_{\alpha}^{
\star}t^{\alpha}]$ becomes
\begin{equation}
P_s(x,t)=\frac{1}{\sqrt{4\pi D_{\alpha}^{\star}t^{\alpha}}\Gamma(1+1/\kappa)}
\int_0^{\infty}\tilde{D}^{-1/2}e^{-\tilde{D}^{\kappa}-\lambda/\tilde{D}}d\tilde{D}.
\label{lapint3}
\end{equation}
After change of variables according to $y=\tilde{D}^{\kappa}$ we have
\begin{eqnarray}
\nonumber
P_s(x,t)&=&\frac{1}{\sqrt{4\pi D_{\alpha}^{\star}t^{\alpha}}\kappa\Gamma(1+1/
\kappa)}\\
&&\times\int_0^{\infty}y^{-1+1/(2\kappa)}e^{-y-\lambda y^{-1/\kappa}}dy.
\label{lapint2}
\end{eqnarray}
With the identification
\begin{equation}
\label{super_stretched}
e^{-z}=H^{1,0}_{0,1}\left[z\left|\begin{array}{l}\rule{1.4cm}{0.02cm}\\(0,1)
\end{array}\right.\right]
\end{equation}
with the Fox $H$-function \cite{mathai}, using the Laplace transform rules for the
$H$-function \cite{wgg} along with the standard
rules for the Fox $H$-function \cite{mathai} one arrives at the result
\begin{eqnarray}
\nonumber
P_s(x,t)&=&\frac{1}{\Gamma(1/\kappa)\sqrt{4\pi D_{\alpha}^{\star}t^{\alpha}}}\\
&&\hspace*{-1.2cm}
\times H^{2,0}_{0,2}\left[\frac{x^2}{4D_{\alpha}^{\star}t^{\alpha}}\left|
\begin{array}{l}\rule{1.4cm}{0.02cm}\\(1/[2\kappa],1/\kappa),(0,1)\end{array}
\right.\right].
\end{eqnarray}

The asymptotic behavior is then \cite{mathai}
\begin{eqnarray}
\nonumber
P_s(x,t)&\simeq&\frac{|x|^{(1-\kappa)/(1+\kappa)}}{\Gamma(1/\kappa)\sqrt{\pi}
(4D_{\alpha}^{\star}t^{\alpha})^{1/(1+\kappa)}}\\
&&\hspace*{-2.4cm}
\times\exp\left(-\frac{1+\kappa}{\kappa^{\kappa/(1+\kappa)}}
\left[\frac{x^2}{4D_{\alpha}^{\star}t^{\alpha}}\right]^{\kappa/(1+\kappa)}
\right).
\label{stretched_as}
\end{eqnarray}

In Figure \ref{fig_stretched} we show the behavior of the resulting probability
density (\ref{stretched_conv}) from numerical inversion. For compressed exponential
distributions $p_D(D_{\alpha})$ with $\kappa>1$ the resulting function $P_s(x,t)$ is
a stretched Gaussian, while for a stretched exponential $p_D(D_{\alpha})$ with $0<
\kappa<1$ the function $P_s(x,t)$ is a superstretched Gaussian, which is broader
than the exponential (Laplace) distribution. Figure \ref{fig_stretched} also
demonstrates that the asymptotic behavior (\ref{stretched_as}) indeed fits the
numerical inversion.

\subsection*{Asymptotics by Laplace's method}

The asymptotic behavior (\ref{stretched_as}) may also be obtained by the Laplace
method. As this is an interesting alternative method to derive the asymptotic
behavior of the integral (see Eq.~(\ref{lapint3}))
\begin{equation}
\label{lapint}
I=\int_0^{\infty}\tilde{D}^{-1/2}e^{-\tilde{D}^{\kappa}-\lambda/\tilde{D}}d\tilde{D}
=\int_0^{\infty}y^{-3/2}e^{-y^{-\kappa}-\lambda y}dy
\end{equation}
for $\lambda\gg1$, we include this approach here. This is a Laplace integral of
the form
\begin{equation}
I=\int_0^{\infty}f(y)e^{-\lambda y}dy.
\end{equation}
The standard methods to evaluate the asymptotics of $I$ cannot be applied, since
$f(y)$ in Eq.~(\ref{lapint}) equals zero at $y=0$ with all its derivatives. Thus,
to evaluate the asymptotics we need to find the maximum of the function
\begin{equation}
\varphi(y)=-\lambda y-y^{-\kappa}
\end{equation}
which is reached at $y_m=(\kappa/\lambda)^{1/(1+\kappa)}$. We now introduce the
new variable $t=y/y_m$, such that Eq.~(\ref{lapint}) becomes
\begin{eqnarray}
\nonumber
I&=&\left(\frac{\lambda}{\kappa}\right)^{1/[2(1+\kappa)]}\int_0^{\infty}t^{-3/2}
\exp\left(-\lambda^{\kappa/(1+\kappa)}\right.\\
&&\left.\times\left[\kappa^{-\kappa/(1+\kappa)}t^{-\kappa}
+t\kappa^{1/(1+\kappa)}\right]\right).
\end{eqnarray}
After substitution $\tau=t\kappa^{1/(1+\kappa)}$ we get
\begin{equation}
\label{lapint1}
I=\lambda^{1/[2(1+\kappa)]}\int_0^{\infty}\tau^{-3/2}e^{\overline{\lambda}S(\tau)}
d\tau,
\end{equation}
where $\overline{\lambda}=\lambda^{\kappa/(1+\kappa)}$ and $S(\tau)=-\tau-\tau^{-
\kappa}$.

Now, the standard Laplace method can be applied to Eq.~(\ref{lapint1}). The function
$S(\tau)$ reaches its maximum at $\tau_m=\kappa^{1/(1+\kappa)}$. Following the
standard procedure we find
\begin{eqnarray}
\nonumber
I&\sim&\lambda^{1/[2(1+\kappa)]}\int_{\tau_m-\varepsilon}^{\tau_m+\varepsilon}
\tau^{-3/2}\\
\nonumber
&&\times\exp\left(\overline{\lambda}\left[S(\tau_m)+\frac{(\tau-\tau_m)^2}{2}
S''(\tau_m)\right]\right)\\
\nonumber
&\sim&\lambda^{1/[2(1+\kappa)]}\tau_m^{-3/2}e^{\overline{\lambda}S(\tau_m)}
\int_{\infty}^{\infty}\exp\left(-\frac{\overline{\lambda}}{2}\left|S''(\tau_m)
\right|\tau^2\right)\\
&=&\lambda^{1/[2(1+\kappa)]}\tau_m^{-3/2}e^{\overline{\lambda}S(\tau_m)}\sqrt{
\frac{2\pi}{\overline{\lambda}S''(\tau_m)}}.
\end{eqnarray}
With $S(\tau_m)=-(1+\kappa)\kappa^{-\kappa/(1+\kappa)}$ and $S''(\tau_m)=-
(1+\kappa)\kappa^{-1/(1+\kappa)}$ and applying this result to the above
probability density function (\ref{lapint3}), we obtain the same asymptotic
behavior (\ref{stretched_as}), up to a numerical prefactor.

\section*{Power law diffusivity distribution}

We now consider the power law distribution
\begin{equation}
p_D(D)=\frac{\alpha D_{\star}^{\alpha}}{(D_{\star}+D)^{1+\alpha}}
\end{equation}
with $\alpha>0$. With the relation (\ref{fouriersuper}) we separately consider
the following cases:

\textbf{(i)} $0<\alpha<1$. The Laplace transform of the diffusivity distribution
reads
\begin{eqnarray}
\nonumber
\tilde{p}_D(s)&=&\alpha D_{\star}^{\alpha}\int_0^{\infty}\frac{dD}{(D_{\star}+D)^{
1+\alpha}}e^{-Ds}dD\\
&&\hspace*{-2.2cm}
=1-D_{\star}s^{\alpha}e^{D_{\star}s}\left(\Gamma(1-\alpha)-\int_0^{D_{\star}s}
z^{-\alpha}e^{-z}dz\right),
\end{eqnarray}
after substituting and integrating by parts. In the tails we then obtain the
following scaling behavior for the probability density function,
\begin{eqnarray}
\nonumber
\lim_{k\to0}P(k,t)&\sim&1-D_{\star}k^{2\alpha}t^{\alpha}\left(1+D_{\star}k^2t+
\ldots\right)\\
&&\hspace*{-2.6cm}
\times\left[\Gamma(1-\alpha)-\frac{(D_{\star}k^2t)^{1-\alpha}}{1-\alpha}+
\frac{(D_{\star}k^2t)^{2-\alpha}}{2-\alpha}-\ldots\right].
\end{eqnarray}
It thus follows that
\begin{equation}
P(x,t)\simeq\frac{1}{|x|^{2\alpha+1}}
\end{equation}
such that the second moment does not exist.

\textbf{(ii)} $\alpha=1$. After integrating by parts once we obtain
\begin{equation}
P(k,t)\sim1-D_{\star}k^2t\log(D_{\star}k^2t).
\end{equation}
The second moment still does not exist.

\textbf{(iii)} $1<\alpha<2$. Integrating by parts twice, we find
\begin{eqnarray}
\nonumber
\tilde{p}_D(s)&=&1-\frac{D_{\star}s}{\alpha-1}+\frac{D_{\star}^{\alpha}s^{\alpha}
e^{D_{\star}s}}{\alpha-1}\\
&&\hspace*{-0.8cm}
\times\left(\Gamma(2-\alpha)-\int_0^{D_{\star}s}z^{1-\alpha}e^{-z}dz\right),
\end{eqnarray}
such that we obtain
\begin{equation}
P(x,t)\simeq\frac{1}{|x|^{2\alpha+1}}
\end{equation}
with the MSD
\begin{equation}
\langle x^2(t)\rangle\sim\frac{2D_{\star}}{\alpha-1}t.
\end{equation}

Power law diffusivity distributions lead to a long tailed, power law distribution
$P(x,t)$ in the superstatistical approach. The second moment diverges for $0<
\alpha\le1$, while normal diffusion emerges for $\alpha>1$.

\section{Dimensionless units for the minimal model}
\label{app_dim}

To simplify the calculations and obtain a more elegant formulation we introduce
dimensionless variables according to $t'=t/t_0$ and $x'=x/x_0$ (and similarly
for the $y$ and $z$ components). For the $x$ component, the set (\ref{lang}) of
stochastic equations then becomes
\begin{subequations}
\label{dimlang}
\begin{eqnarray}
\frac{d}{dt'}x'(t)&=&\frac{t_0}{x_0}\sqrt{2D(t)}\xi(t_0t')\\
D(t)&=&Y^2(t)\\
\frac{d}{dt'}Y&=&-\frac{Y}{\tau/t_0}+\sigma\eta(t_0t').
\end{eqnarray}
\end{subequations}
Noting that for the Gaussian noise sources we have $\xi(t_0t')=t_0^{-1/2}\xi(t')$
and $\eta(t_0t')=t_0^{-1/2}\eta(t')$ we rewrite Eqs.~(\ref{dimlang}) as
\begin{subequations}
\begin{eqnarray}
\frac{d}{dt'}x'(t)&=&\sqrt{2\overline{D}(t)}\xi(t')\\
\overline{D}(t)&=&\overline{Y}^2(t)\\
\frac{d}{dt'}\overline{Y}&=&-\frac{\overline{Y}}{\overline{\tau}}+\overline{\sigma}
\eta(t'),
\end{eqnarray}
\end{subequations}
where
\begin{equation}
\overline{D}=\frac{t_0}{x_0^2}D,\quad \overline{Y}=\frac{t_0^{1/2}}{x_0}Y,
\quad \overline{\tau}=\frac{\tau}{t_0}.
\end{equation}
Now we choose the temporal and spatial scales such that $\overline{\tau}=
\overline{\sigma}=1$, that is,
\begin{equation}
t_0=\tau,\quad x_0=\sigma\tau.
\end{equation}
With this choice of units the stochastic equations of our minimal diffusing
diffusivity model are then given by Eqs.~(\ref{dimlesslang}).

\section{Two examples for the process $\mathbf{Y}(t)$}
\label{appb}

\subsection{The case $d=2$ and $n=1$}

As an example for the case when the dimensionality of the process $\mathbf{Y}(t)$
differs
from the embedding dimension $d$ of the process $\mathbf{r}(t)$ we take the case
with $d=2$ and $n=1$. In the short time limit we get from Eq.~(\ref{fouprop})
that
\begin{eqnarray}
\nonumber
P(\mathbf{r},t)&\sim&\frac{1}{2\pi t^{1/2}}\int_0^{\infty}\frac{kJ_0(kr)}{\sqrt{
k^2+1/t}}dk\\
&=&\frac{1}{\sqrt{2\pi^3}t}\left(\frac{r}{\sqrt{t}}\right)^{-1/2}K_{1/2}
\left(\frac{r}{\sqrt{t}}\right).
\label{altpdf}
\end{eqnarray}
The asymptotic behavior is given by
\begin{equation}
P(\mathbf{r},t)\sim\frac{1}{2\pi r\sqrt{t}}e^{-t/\sqrt{t}}.
\end{equation}
Comparing this result with Eq.~(\ref{short2}) for the case $d=n=2$ we recognize
the modified prefactor, including a different scaling in $r$ and $t$. Thus the
difference in the dimensionality of the process $\mathbf{Y}(t)$ does not change the
dominating exponential behavior.

The connection to the superstatistical approach in analogy to the discussion in
Section \ref{secsuperstat} following Eq.~(\ref{super}) with the two-dimensional
Gaussian kernel
\begin{equation}
G(\mathbf{r},t|D)=\frac{1}{4\pi D t}e^{-r^2/(4Dt)}
\end{equation}
and the stationary diffusivity distribution
\begin{equation}
p_D^{\mathrm{st}}(D)=\frac{1}{\sqrt{\pi DD_{\star}}}e^{-D/D_{\star}}
\end{equation}
for the case $n=1$ produces the distribution
\begin{equation}
P(\mathbf{r},t)=\frac{1}{\sqrt{2\pi^3}D_{\star}t}\left(\frac{r}{\sqrt{D_{\star}t}}
\right)^{-1/2}K_{1/2}\left(\frac{r}{\sqrt{D_{\star}t}}\right).
\end{equation}
which matches exactly Eq.~(\ref{altpdf}) written in dimensional form.

\subsection{Infinite-dimensional process $\mathbf{Y}(t)$}

We here consider the limit of large dimension $n$ for the process $\mathbf{Y}(t)$.
For short times $t\ll1$ the tails of the probability density $P(\mathbf{r},t)$
follow from (compare Eq.~(\ref{propfou}))
\begin{eqnarray}
\nonumber
\hat{P}(\mathbf{k},t)&\sim&\frac{e^{nt/2}}{\left[\frac{t}{2}\left(1+2k^2+1\right)
+1\right]^{n/2}}\\
\nonumber
&=&\frac{e^{nt/2}}{\left[1+\frac{t(1+k^2)\frac{n}{2}}{n/2}\right]^{
n/2}}\\
&\to&\frac{e^{nt/2}}{e^{(1+k^2)nt/2}}=e^{-ntk^2/2}.
\end{eqnarray}
Thus the tails of the probability density function $P(x,t)$ are Gaussian already
at short times.

The long time behavior $t\gg1$ leads to
\begin{equation}
\hat{P}(\mathbf{k},t)\sim\frac{2^{n/2}\exp\left(\frac{nt}{2}\left[1-\sqrt{1+2k^2}
\right]\right)}{\left[1+\frac{1}{2}\left(\sqrt{1+2k^2}+\frac{1}{\sqrt{1+2k^2}}
\right)\right]^{n/2}}.
\end{equation}
Considering the tails, we take $k\ll1$, revealing that
\begin{equation}
\hat{P}(\mathbf{k},t)\sim\exp\left(-\frac{nk^2t}{2}\right)
\end{equation}
is also Gaussian, with the same variance. Thus, in the high-dimensional case the
regime of exponential wings in the probability density function does not exist at
all and the Gaussian shape is established early on. This is equivalent to the
observation that the kurtosis (\ref{kurt}) becomes Gaussian already for short
times when $n$ is large.

\section{Fourth moment of $P(x,t)$}
\label{appa}

By help of relation (\ref{fourierp}) we obtain the fourth moment of $P(x,t)$.
The necessary derivative of $\tilde{T}_n(s,t)$ with respect to $s$ is
\begin{widetext}
\begin{eqnarray}
\nonumber
\frac{\partial\tilde{T_n}(s,t)}{\partial s}&=&-\frac{ne^{nt/2}}{2}
\left[\frac{1}{2}\left(\sqrt{1+2s}+\frac{1}{\sqrt{1+2s}}\right)\sinh
\left(t\sqrt{1+2s}\right)
+\cosh\left(t\sqrt{1+2s}\right)\right]^{-n/2-1}\left[\frac{s}{(1+2s)^{3/2}}
\sinh\left(t\sqrt{1+2s}\right)\right.\\
&&+\frac{t}{2}\left(1+\frac{1}{1+2s}\right)\cosh\left(t\sqrt{1+2s}
+\frac{t}{\sqrt{1+2s}}\sinh\left(t\sqrt{1+2s}\right)\right],
\end{eqnarray}
\end{widetext}
and thus
\begin{equation}
\left.\frac{\partial\tilde{T}_n(s,t)}{\partial s}\right|_{s=0}=-\frac{nt}{2}.
\end{equation}
The second differentiation and subsequent limit $s\to0$ produces, after some
steps,
\begin{equation}
\frac{\partial^2\tilde{T_n}(s,t)}{\partial s^2}\Big|_{s=0}=\left<\tau^2(t)\right>
=-\frac{n}{4}\left(1-e^{-2t}\right)+\frac{nt}{2}+\frac{n^2t^2}{4}.
\end{equation}

\acknowledgments

FS and AVC acknowledge Amos Maritan and Samir Suweis for stimulating discussions.
AVC and RM acknowledge funding from the Deutsche Forschungsgemeinschaft.
AVC acknowledges financial support from Deutscher Akademischer Austauschdienst
(DAAD).

\end{document}